\documentclass[aps, pre, preprint, superscriptaddress, longbibliography, nobalancelastpage]{revtex4-1}
\usepackage{graphicx}
\usepackage{amssymb}
\usepackage{amsmath}
\usepackage{hyperref}
\usepackage{float}
\usepackage{mathtools}
\usepackage{bm}
\usepackage{dcolumn}
\usepackage[dvipsnames]{xcolor}
\begin{document}
\title{Mechanical properties of DNA and DNA nanostructures: comparison of atomistic, Martini and oxDNA}

\author{Supriyo Naskar}
\email{supriyo@iisc.ac.in}
\affiliation{Center for Condensed Matter Theory, Department of Physics, Indian Institute of Science, Bangalore 560012, India}
\author{Prabal K. Maiti}
\thanks{Corresponding author}
\email{maiti@iisc.ac.in}
\affiliation{Center for Condensed Matter Theory, Department of Physics, Indian Institute of Science, Bangalore 560012, India}
\date{\today}
\begin{abstract}
The flexibility and stiffness of small DNA play a fundamental role ranging from several biophysical processes to nano-technological applications. Here, we estimate the mechanical properties of short double-stranded DNA (dsDNA) having length ranging from 12 base-pairs (bps) to 56 bps, paranemic crossover (PX) DNA, and hexagonal DNA nanotubes (DNTs) using two widely used coarse-grain models -- Martini and oxDNA. To calculate the persistence length ($L_p$) and the stretch modulus ($\gamma$) of the dsDNA, we incorporate the worm-like chain and elastic rod model, while for DNT, we implement our previously developed theoretical framework. We compare and contrast all the results with previously reported all-atom molecular dynamics (MD) simulation and experimental results. The mechanical properties of dsDNA ($L_p$ $\sim$ 50nm, $\gamma$ $\sim$ 800-1500 pN), PX DNA ($\gamma$ $\sim$ 1600-2000 pN) and DNTs ($L_p$ $\sim$ 1-10 $\mu$m, $\gamma$ $\sim$ 6000-8000 pN) estimated using Martini soft elastic network and oxDNA are in very good agreement with the all-atom MD and experimental values, while the stiff elastic network Martini reproduces order of magnitude higher values of $L_p$ and $\gamma$. The high flexibility of small dsDNA is also depicted in our calculations. However, Martini models proved inadequate to capture the salt concentration effects on the mechanical properties with increasing salt molarity. oxDNA captures the salt concentration effect on small dsDNA mechanics. But it is found to be ineffective to reproduce the salt-dependent mechanical properties of DNTs. Also, unlike Martini, the time evolved PX DNA and DNT structures from the oxDNA models are comparable to the all-atom MD simulated structures. Our findings provide a route to study the mechanical properties of DNA and DNA based nanostructures with increased time and length scales and has a remarkable implication in the context of DNA nanotechnology.
\end{abstract}

\maketitle
\section{Introduction}
DNA, the genetic information carrier of life, having contour length of few cm to meters, has to undergone twisting, bending, stretching to be packaged into a cell of few $\mu$m diameter\cite{60}. To facilitate such packaging into the cell, DNA is wrapped around oppositely charged histone proteins in a highly organized structure known as nucleosome core particle (NCP)\cite{62}. Several other fundamental biological processes such as transcriptions, replications, gene expression are also bringing in mechanical stress to DNA\cite{13, 36, 56, 57}. In most of these biological processes, the length scales of DNA involved are of few tens of base pairs and much smaller than its persistence length ($\sim$50 nm)\cite{5, 70, 74}. Hence, understanding the mechanics of small DNA is of key interest in the field of molecular biology.\par

Sequence specificity of DNA has made it possible to utilize it as an ideal building block to create a wide variety of complex nanostructures\cite{68}. Utilizing the key properties of DNA such as its persistence length and Watson-Crick sequence specificity, nanometer-scale devices and material can be created. The idea of the DNA nanotechnology field was first led by N. C. Seeman in 1982 who designed a four-armed junction, a cube, a truncated octahedron in several stages\cite{15, 16, 27, 68, 90}. Seeman’s visionary idea is now rapidly developing the field of DNA nanotechnology because of its potential application in drug delivery, synthetic biology and DNA computing\cite{7, 30, 31, 37, 38, 59, 68, 69, 82}. In another approach namely ‘DNA origami’, recently established by P. Rothemund,  a long DNA strand can be folded using small DNA staple strands and can be very efficient in designing complex DNA nanostructures\cite{64}. DNA nanotube (DNT), a programable biomimicking ion-channel, is one of the recent additions of the expanding repository of DNA nanostructures\cite{24, 29, 32, 65, 83}. Once properly derivatized, DNT can spontaneously insert into the lipid membrane and have appeared to function as an ion channel and can also transport drug molecules across the cell membranes\cite{9,10,11, 18, 22, 25, 26, 49, 87}. Also, DNT can be operated as a robotic arm to carry and transport cargo across the cell and can also perform as a support system for other cell membrane channels\cite{28}. Enhanced Rigidity of DNTs have widespread applications from nanoelectronics to nanomechanical devices\cite{29, 33, 35, 83, 88}. Thus, the mechanical strength of the DNTs is one of the most essential aspects and considerable effort has been put to measure and increase the mechanical strength of these nanotubes.\par

During the last few decades, with the advancement of single-molecule manipulation techniques, our understanding of DNA and DNA nanostructure’s mechanics has expanded immensely\cite{1, 5, 12, 13, 20, 73, 74, 85}. Utilizing experimental techniques such as optical trapping, magnetic tweezers, atomic force microscopy (AFM), it has now turn out to be feasible to externally pull DNA at different physiological conditions\cite{12, 41, 67, 73, 74}. Although the mechanics of DNA is highly dependent on the sequence and physiological condition, most of these pulling experiments are suitably described by a worm-like chain (WLC) model\cite{12}. AFM studies by Mazur et al. observed that the WLC model can properly capture the flexibility of short DNA beyond two helical turns\cite{44, 45}. However, some experimental studies challenged the validity of the WLC model for very short DNA and at high force regimes. Modifications have been made to the WLC model (such as twistable worm-like chain model, linear sub elastic chain model) to explain the mechanics of DNA at high force regimes\cite{19, 53, 61}. Likewise, to obtain the mechanical properties of DNTs, the contour length and end-to-end distances have been extracted from fluorescence images and then fitted to the WLC model to estimate their values of persistence length\cite{83}. In a recent study by Liu et. al, combining small-angle Xray scattering (SAXS) and Förster resonance energy transfer (FRET) characterization with MD simulation, estimated structural and mechanical properties of DNA in a variety of physiological conditions\cite{34}. \par
Computational modeling like molecular dynamics or density functional theory can play a key role given the fact that sometimes experiments are very difficult to perform routinely. Also, it is challenging to obtain high-resolution structural information from experiments. Theoretical models and all-atom simulations can be very effective in predicting structural and mechanical properties prior to the experimental realizations\cite{14, 37, 38, 52, 97, 72, 91, 92, 95, 96}. Notably, Maiti and coworkers implemented the WLC and elastic rod model to study the mechanics of dsDNA and DNA based nanostructures like paranemic crossover (PX/JX) DNA molecules, DNA nanotubes, and found their elastic properties to be in close agreement with the experimental results\cite{17, 24, 25, 26, 48, 49, 66}.\par

Simulating fully atomistic models of dsDNA and DNA nanostructures are computationally very demanding. A less computationally expensive and highly efficient approach is the coarse-grained modeling in which the basic units are no longer single atom, but some larger units -- be it some atoms, or a nucleotide, a base pair or a double helix\cite{3, 47, 55, 75, 76, 78, 79}. Coarse-grained simulations averages over the nonessential degrees of freedom and thus reduces the complexity of the fully atomistic simulations. Therefore, such methods inevitably lose structural details of the systems and accuracy of the physical properties depend on the level of parameterization but can be very effective in studying system over longer time and length scales. In this study, we have used two widely used coarse-grained DNA models -- Martini and oxDNA to understand the mechanics of short DNAs and DNA based nanostructures. The organization of the paper is as follows: In the method section, we give a brief description of the models followed by the simulation methodology. We then provide the theory used in this study to estimate the mechanical properties of DNAs and DNTs. In the result section, we present our calculation of stretch modulus and persistence lengths of dsDNA having various lengths, PX DNA and DNT and compare those with all-atom simulation results. Finally, we conclude and provide appealing future directions. \par

\section{Methods}
\subsection{Martini model}

The coarse-grained (CG) Martini model of DNA is parameterized by combining top-down information derived from the experiment and bottom-up knowledge gained from the atomistic simulations. For each residue, the beads are divided into side-chain beads (SC1, SC2, SC3, SC4) and backbone beads (BB1, BB2, BB3). The backbone of the DNA is modeled with three beads -- one for phosphate group (BB1 bead) and the other two represents the sugar moiety (BB2 and BB3). Roughly four non-hydrogen atoms are mapped to one CG bead. The pyrimidines (C and T) are mapped as three beads and the purines (A and G) are mapped as four beads. SC1 and SC2 beads form base-pair when they are in opposite strands. The CG representation of Martini model is illustrated in figure \ref{fig1}a. The backbone beads are modeled with a large sphere with diameter  $\sigma$= 0.47 nm (for sugar beads $\sigma$= 0.43 nm). Since the bases are stacked very closely to each other with a distance of 0.34 nm, these CG beads with such high $\sigma$ will lead to overlapping with each other. For this reason and also to maintain the planarity of the bases, a small CG bead with $\sigma$= 0.32 nm is parameterized to characterize the bases. The interaction potential, $V$ is composed of bonded interaction, $V_{bonded}$ and non-bonded interactions, $V_{non-bonded}$. The bonded potential is composed of usual harmonic bond, angle and dihedral potential. The bonded interactions are modeled in a manner that reproduces bond, angle and dihedral distributions obtained from the atomistic MD simulations. The non-bonded potential is composed of Columbic interaction between charge particles and Van der Waals interaction. The bead types are selected by comparing the calculated partitioning free energies of DNAs in water, chloroform and hydrated octanol with available experimental data. To incorporate the hydrogen bond interactions between the nucleobases, the interaction is tuned between the hydrogen-bonded beads. An elastic network has been implemented in order to maintain the canonical form of the dsDNA. All the beads are chargeless expect the one representing the phosphate group, which has been assigned with -1e charge. Like all CG models, Martini also has its advantages and disadvantages. The partitioning free-energy of nucleobases in the Martini model is comparable to the experimental values. In some cases, it reproduces better results than the all-atom force-fields. The radius of gyration of ssDNA and ion distribution around DNA are also reproduced reasonably well compared to those obtained at the all-atom level. The two different types of elastic networks serve different purposes and the choice of the network will depend on the particular need in the application. The current limitation of the model is the reduced strength of the base-pairing interactions. Also, the hydrogen bonds between the base pairs do not have depends on the angle, which prevents this model from studying  DNA hybridization, melting, hairpin formation, or intercalation. For a detailed description of the Martini model, the reader is recommended to go through the original references \cite{42, 78, 79}. \par
\subsection{OxDNA model}
\begin{figure}[htbp]
 \centering
 \includegraphics[width=6.45 in,keepaspectratio=true]{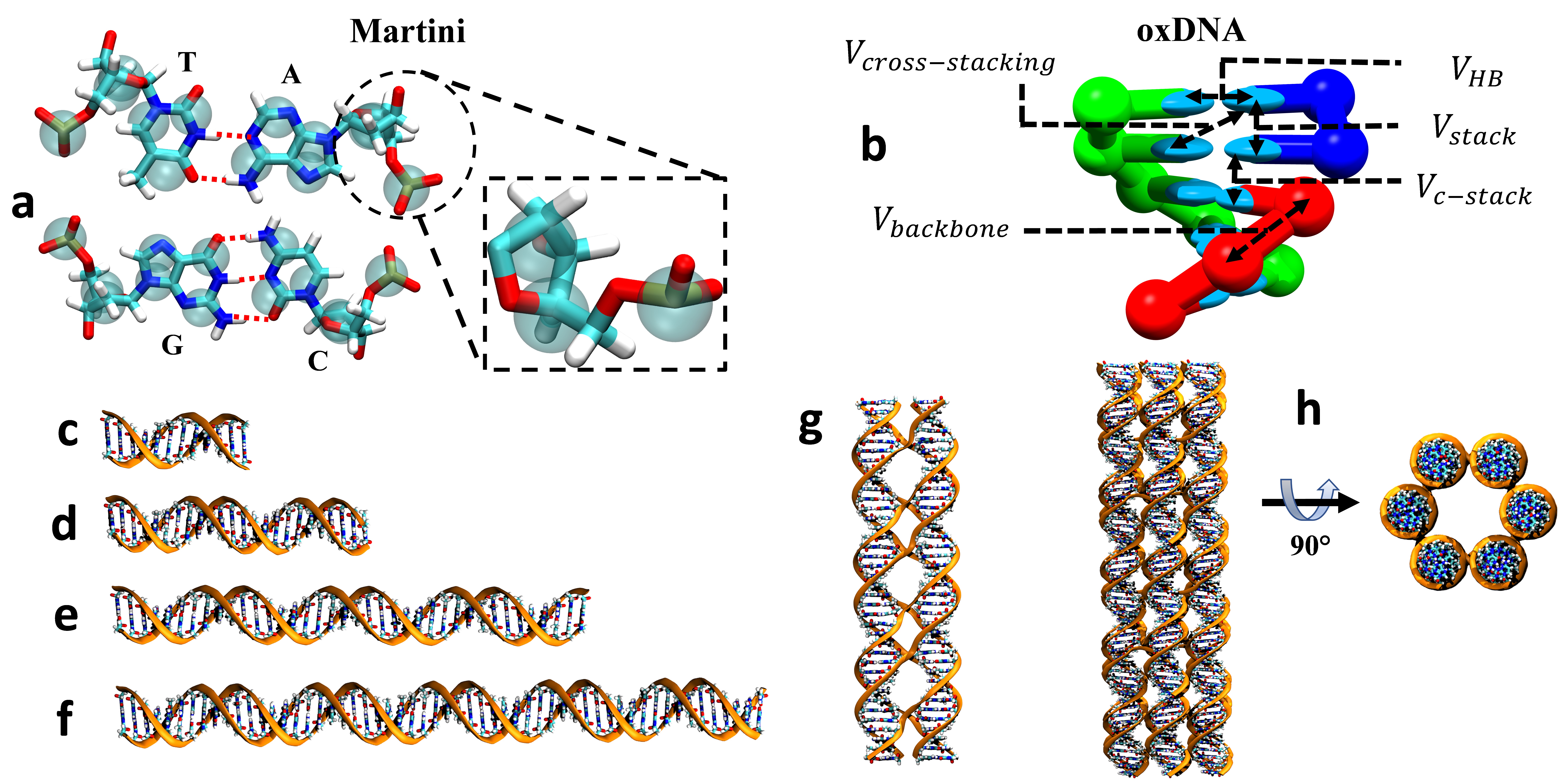}
 \caption{Description of the coarse-grained models and various systems studied in this work. (a) Coarse-grain mapping of Martini DNA. Each base of the DNA is modeled with either three beads (Cytosine and Thymine) or four beads (Adenine and Guanine). The sugar moiety is modeled with two beads and the phosphate group is modeled with a single bead. (b) Coarse-grained oxDNA model. Each nucleotide is modeled with two beads – one for the base (cyan ellipsoid) and another for the backbone (red and blue sphere) of the DNA. Initial structures of dsDNA (c) 12bps (d) 24 bps (e) 38 bps (f) 56 bps. (g) Initial structure of paranemic crossover DNA. (h) Initial model of hexagonal DNA nanotube.}
 \label{fig1}
\end{figure}

The CG oxDNA model is parameterized with a top-down approach, aimed to include the thermodynamic transition of DNA\cite{55, 75, 76}. Each strand of dsDNA is constructed using a chain of CG beads. Each nucleobase is represented by a single CG bead as illustrated in figure \ref{fig1}b. For the phosphate backbone, another CG bead is assigned (See Figure \ref{fig1}b). There are three interaction sites for each nucleotide and they lie in a straight line. The three interaction sites are the hydrogen-bonding site, base stacking site and backbone excluded volume site. A normal vector is defined to specify the orientational plane of the bases. Each CG bead is interacting with others in a pairwise fashion with excluded volume, stacking, hydrogen-bonding (HB), cross-stacking interaction and co-axial stacking (c-stack). Nearest neighbors (nn) interactions are distinct compared to all other interactions in the strand which permits for stacking and strand connectivity. The potential energy function of the oxDNA model can be written as a sum of the following interactions, 
\begin{equation}
V= \sum_{nn}  (V_{backbone} + V_{stack}+V_{exc}^{'} ) +\sum_{other  pairs} (V_{HB}+V_{cross-stacking}+V_{exc}+V_{c-stack} )  \label{1}
\end{equation}
The nearest neighbor pairs are interacting via $V_{backbone}$, $V_{stack}$  and  $V_{exc}^{'}$  potentials and the other pairs are interacting with $V_{HB}$, $V_{cross-stacking}$, $V_{exc}$  and $V_{c-stack}$  potentials. $V_{backbone}$ represents the covalent bond between the rigid bodies in the form of a finitely extensible nonlinear elastic (FENE) spring with an equilibrium distance of 6.4 $\AA$ to hold the nucleotides in a strand together. $V_{stack}$ is a smoothly cut-off Morse potential with a minimum at 3.4 $\AA$ to mimic the stacking of the bases in a coplanar manner. $V_{exc}^{'}$ and $V_{exc}$ are the excluded volume interactions between nearest neighbor and next-nearest neighbor, respectively. It provide the stiffness to unstacked bases of the single strands and prevents crossing of the chains. $V_{HB}$ is a smoothly cut-off Morse potential between hydrogen-bonding sites and provides the hydrogen-bonded interaction for base-pairing. $V_{cross-stacking}$ and $V_{c-stack}$ corresponding to the cross-stacking and co-axial stacking between the non-nearest neighbors, respectively and provides additional stabilization of the duplex. In the oxDNA2 model, which is a revised version of the original oxDNA model, sequence-dependent stacking and hydrogen-bonding interactions are used \cite{76}. No explicit electrostatic interaction is taken into account in the model. Too include electrostatic effect, Debye-H{\"u}ckel potential is added in the non-bonded interaction. 
OxDNA has several advantages over the other CG models. This is the first CG model which captures the three key thermodynamic processes that affect self-assembly-- hybridization of double-stranded DNA, single-stranded stacking, hairpin formation. The mechanical properties like stretch modulus, torsional modulus and persistence length are also reasonably well represented by the oxDNA model. The only limitation of this model that it under predict the effect of terminal mismatches, dangling ends, bulges and internal mismatches on the duplex melting temperature. For a detailed description of the oxDNA model, the reader is recommended  to the original the references \cite{55, 75, 76}.

\subsection{Simulation procedure}
Here, we have studied four dsDNA systems of finite length, 12, 24, 38, and 56 bps, PX DNA, and a hexagonal DNA nanotube originally synthesized by Wang et. al (See figure \ref{fig1}c-h). The sequence of dsDNA, PX DNA and the DNT is provided in the supporting information (SI). The all-atom structures of the dsDNA and the DNT are built using a custom written code in nucleic acid builder (NAB) language of Ambertools. Then the structures are converted to the CG Martini model using Martinize.py code\cite{78}. For each system, two different elastic networks -- soft and stiff has been implemented. The structures are then solvated in CG Martini water model and an appropriate number of sodium ion (Na$^+$) is added to make the whole system charge neutral. For Martini simulations, we have used two different types of water models in our study --one is standard chargeless Martini CG water model and other one is Martini polarizable water model. We present all the main results for standard Martini CG water model. We have separately discussed the effect of Martini polarizable water model on DNA mechanics later.\par
The MD simulations of the CG Martini models were performed using Gromacs 5.1 simulation package\cite{81}.  All the systems were subjected to the steepest descent energy minimization while keeping restraint on DNA/DNT with a force constant of 20 kJ.mol$^{-1}$\AA$^{-2}$ which was gradually reduced to zero. The systems were then heated to 300 K and equilibrated in the NPT ensemble for 100 ns. The systems were then subjected to 2 $\mu$s long simulation in the NVT ensemble. The time step was taken to be 10 fs for all the simulations. Long-range electrostatic interactions were included using reaction-field with a cutoff of 1.5 nm\cite{4}. Same cutoff was also used for the short-range part of the Coulomb interaction with the potential shift cutoff scheme. For polarizable water model we have used particle mesh Ewald (PME) method for Coulomb interaction with a cutoff of 1.5 nm. The pair-list was updated after every 10th steps. Temperature regulation was achieved by velocity rescaling thermostat with a time constant of 0.5 ps\cite{81}. To maintain the pressure at 1.0 bar, Berendsen barostat was used with a time constant of 3.0 ps and a compressibility of $3.0 \times 10^{-5}$ bar$^{-1}$\cite{6}.\par 
For oxDNA models, the structures are directly built using the source code\cite{55, 75, 76}. MD simulations with the oxDNA model were performed using the source code provided by with oxDNA force field\cite{55, 75, 76}. We have simulated the systems for $2\times10^8$ to $2\times10^9$ MD cycle with a time step of 0.005 which in real unit corresponds to a total of 3 $\mu$s to 30 $\mu$s long MD run. Simulations were performed in the NVT ensemble at T = 300 K. Temperature regulation was implemented using Anderson thermostat\cite{2}.\par

Analysis of the mechanical properties was done for the last half of the production run. We divided the trajectory equally into 5 segments and for each segment we independently calculated all the relevant properties and finally averaged it over the 5 segments to report the final quantities. Visualization of the systems and trajectories were done using VMD\cite{23} and UCSF Chimera\cite{58}. For trajectory analysis purpose we have used CPPTRAJ\cite{63}, MDTRAJ\cite{46}, home written python codes and TCL scripts.
 \subsection{Theory of dsDNA, PX DNA, and DNT mechanics}
\begin{figure}[htbp]
 \centering
 \includegraphics[width=1.65 in,keepaspectratio=true]{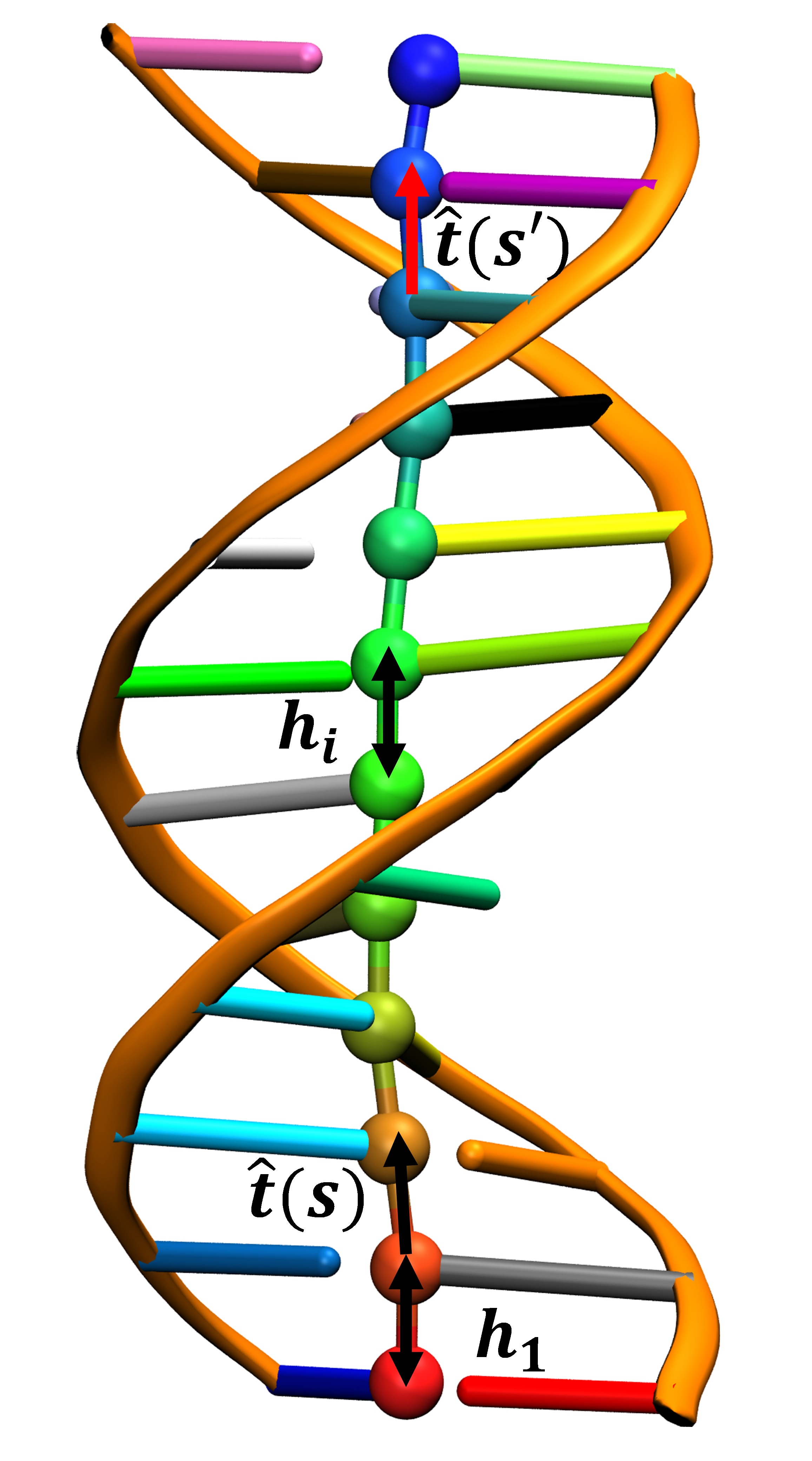}
 \caption{Definition of the contour length and bending angle used in our studies. The spheres represent the center of mass of each base pair. The contour length is calculated by adding the center of mass between two consecutive base pairs as $L=\ \sum_{i=1}^{N}h_i $, where $N$ is the number of base-pairs of the dsDNA. The bending angle is defined as angle between the local vectors $\hat{t}(s)$ and $\hat{t}(s^{'})$ defining the local axial direction of dsDNA such as, $\theta=\cos^{-1}{(\hat{t}\left(s\right).\hat{t}(s^\prime))}$.}
 \label{new_slide}
\end{figure}

 We estimated the dsDNA stretch modulus by combining microscopic elastic rod theory with the WLC model. We assumed dsDNA as a uniform elastic rod of instantaneous contour length, $L$ at time $t$, which is fluctuating about its equilibrium length $L_0$ due to the thermal energy. The definition of dsDNA contour length is given in figure \ref{new_slide}. Any fluctuation of length will give rise to a restoring force (F) which is proportional to ($L-L_0$) and can be expressed as, $F (L)= -\gamma\ (L-L_0)/L_0$, where $\gamma$ is the stretch modulus of the rod. The energy associated with this force can be obtained by integrating the force with respect to $L$ which gives $E(L) = -\gamma\ (L-L_0)^2/(2L_0 )$. The probability $(P (L))$ of having a rod of length, $L$ with an energy $E (L)$ is then obtained through Boltzmann statistics\cite{17, 25, 48, 49},
 
 \begin{equation}
 P(L)=\sqrt\frac{\gamma\beta}{2\pi\ L_0\ }\ \ exp\left[-\frac{\gamma\beta\ L_0}{2}\ \left(\frac{L}{L_0}\ -1\right)^2\ \right]
  \label{2}
\end{equation}

Where $\beta=\frac{1}{K_BT}\ $. From the slope of the straight line of $\ln{P\left(L\right)}$ vs $\left(\frac{L}{L_0}-1\right)^2$, the stretch modulus has been extracted. \par

In order to calculate the bending persistence length $L_P$ of dsDNA, we first estimate the distribution of the bending angle ($\theta$). The $\theta$ is defined as the  angle between the local vectors $\hat{t}(s)$ and $\hat{t}(s^\prime)$ defining the local axial direction of dsDNA or DNT such as $\theta=\cos^{-1}{(\hat{t}\left(s\right).\hat{t}(s^\prime))}$ , where $s$ parameterize the path of the polymer as $s\in(0,L_0)$ (see figure \ref{new_slide}). Similar to the length fluctuation, the probability distribution of $\theta$ can be expressed as a Gaussian distribution as\cite{17, 25, 48, 49},

\begin{equation}
P\left(\theta\right)=\sqrt{\frac{\beta\kappa}{{2\pi L}_0}}exp\left[-\frac{\beta\kappa}{{2L}_0}\theta^2\right]  \label{3}
\end{equation}
\begin{equation}
\Rightarrow\ln{P\left(\theta\right)=}-\frac{L_P}{L_0}\left(1-\cos{\theta}\right)+\frac{1}{2}\ln{\frac{\beta\kappa}{2\pi L_0}}  \label{4}
\end{equation}

Where $L_P=\beta\kappa$ and $\kappa$ is the bending modulus. \par

One can also estimate the persistence length from the WLC model. In the WLC model the mean square end-to-end distance can be expressed as\cite{17, 48},

\begin{equation}
\langle\ R^2\ \rangle = \langle\ \vec{R}.\vec{R}\ \rangle = \langle\ \int_{0}^{L_0} \hat{t}\left(s\right) ds \int_{0}^{L_0} \hat{t}(s^\prime) ds^\prime\ \rangle \label{5a} 
\end{equation}

\begin{equation}
\Rightarrow \langle\ R^2\ \rangle = \int_{0}^{L_0}  ds \int_{0}^{L_0} \langle\ \hat{t}\left(s\right)\ .\ \hat{t}(s^\prime) \ \rangle ds^\prime\ \label{5} 
\end{equation}

\begin{equation}
\Rightarrow \langle\ R^2\ \rangle = \int_{0}^{L_0}  ds \int_{0}^{L_0} exp\left[-\frac{|s-s^\prime|}{L_P}\right]\ ds^\prime\ \label{6} 
\end{equation}

\begin{equation}
\Rightarrow \langle\ R^2\ \rangle=2L_P L_0 \left[ 1-\frac{L_P}{L_0}  \left\{1- exp\left(-\frac{L_0}{L_P}\right)   \right\} \right] \label{7}
\end{equation}

Equation \ref{7} can be solved numerically to get the persistence length , $L_P$. We referred the persistence length calculated using equation \ref{7} as $L_P^{WLC}$. \par

Furthermore, according to the microscopic elastic theory, for any uniform rod of cross-sectional area A and stretch modulus $\gamma$, Young’s modulus $\left(Y\right)$ is expressed as, $Y\ =\frac{\gamma}{A}$ and the bending modulus is given by, $\kappa=IY$ , where $I$ is the area moment of inertia of the rod. Combining the above relation, we can write the stretch modulus as\cite{17, 25, 49}, 

\begin{equation}
\gamma=\frac{A\kappa}{I}=\frac{AL_P}{\beta I} \label{8}
\end{equation}
We assume the dsDNA as a uniform rod which gives the area moment of inertia, $I=\frac{\pi R_1^4}{4}$, $R_1$ being the radius of the dsDNA. For DNT we explicitly calculate the I as\cite{25, 49},
\begin{equation}
I=\frac{1}{2}\left[\sum_{6}{I_0+\left(4R_1^2+2R_2^2\right)A_1+I_0\left(16\frac{R_2^2}{R_1^2}-10\right)}\right] \label{9}
\end{equation}

$R_2$ is the average radius of the DNT. The detail calculation of the $I$ of DNT is given in reference \cite{25}. 

\section{Results and Discussions}
\subsection{Mechanical properties of dsDNA }
\begin{figure}[b!]
 \centering
 \includegraphics[width=6.45in,keepaspectratio=true]{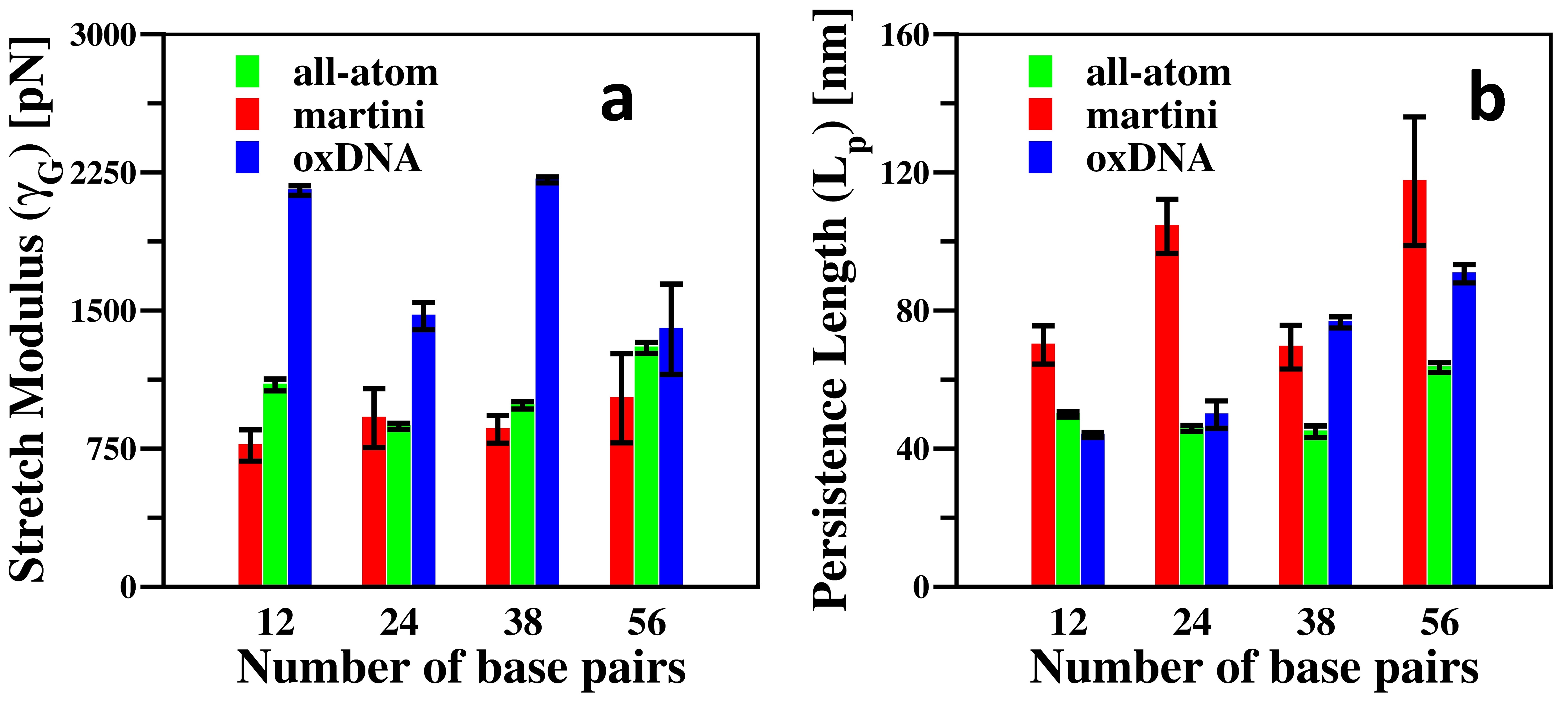}
 \caption{Stretch modulus and persistence length of CG dsDNA having different lengths. The quantities are calculated using equation \ref{2} and \ref{5}. }
 \label{fig2}
\end{figure}
To measure the mechanical properties as mentioned above, we first calculate the contour length distribution and the bending angle distribution of the dsDNA. The contour length is calculated by adding the center of mass of each base pair as shown in figure \ref{new_slide}. In order to calculate the bending angle, two local vectors are defined by joining the center of mass of two consecutive base pairs (See figure \ref{new_slide}). The end base pairs are not considered as they might fray during the simulation. The angle between the two vectors is defined as the bending angle. In figure S5 of SI we have plotted the contour length and bending angle distribution of a 12 bps dsDNA. By fitting the contour length and bending angle distribution using equation \ref{2} and \ref{4} respectively, we have extracted the stretch modulus ($\gamma$) and persistence length ($L_p$) (see figure \ref{fig2} and table \ref{cgtab1}). We find that the persistence lengths	of the oxDNA model are in close agreement with the values obtained from  all atom simulations and match well with experimental values\cite{8, 17, 21, 39, 40, 77, 89}. The persistence lengths estimated from Martini model with soft elastic network are in the range of 60 to 130 nm and slightly higher than the values obtained using all-atom model and experimental values of $L_P$ ($\sim$ 50nm). Additional softening of the elastic network can improve the results but that may lead to numerically unstable simulations. The stretch modulus obtained  from soft-Martini and oxDNA models are in very good agreement with the all-atom and literature values\cite{17, 40}. Furthermore, the CG Martini with stiff elastic network model gives the persistence length and stretch modulus which is an order of magnitude higher (see table S3 of SI). In stiff-Martini, since the force constant of the elastic network is very high, the structures remain highly static and do not exhibit much  fluctuation (see figure S3 of SI).\par
\begin{figure}[htbp]
 \centering
 \includegraphics[width=6.25in,keepaspectratio=true]{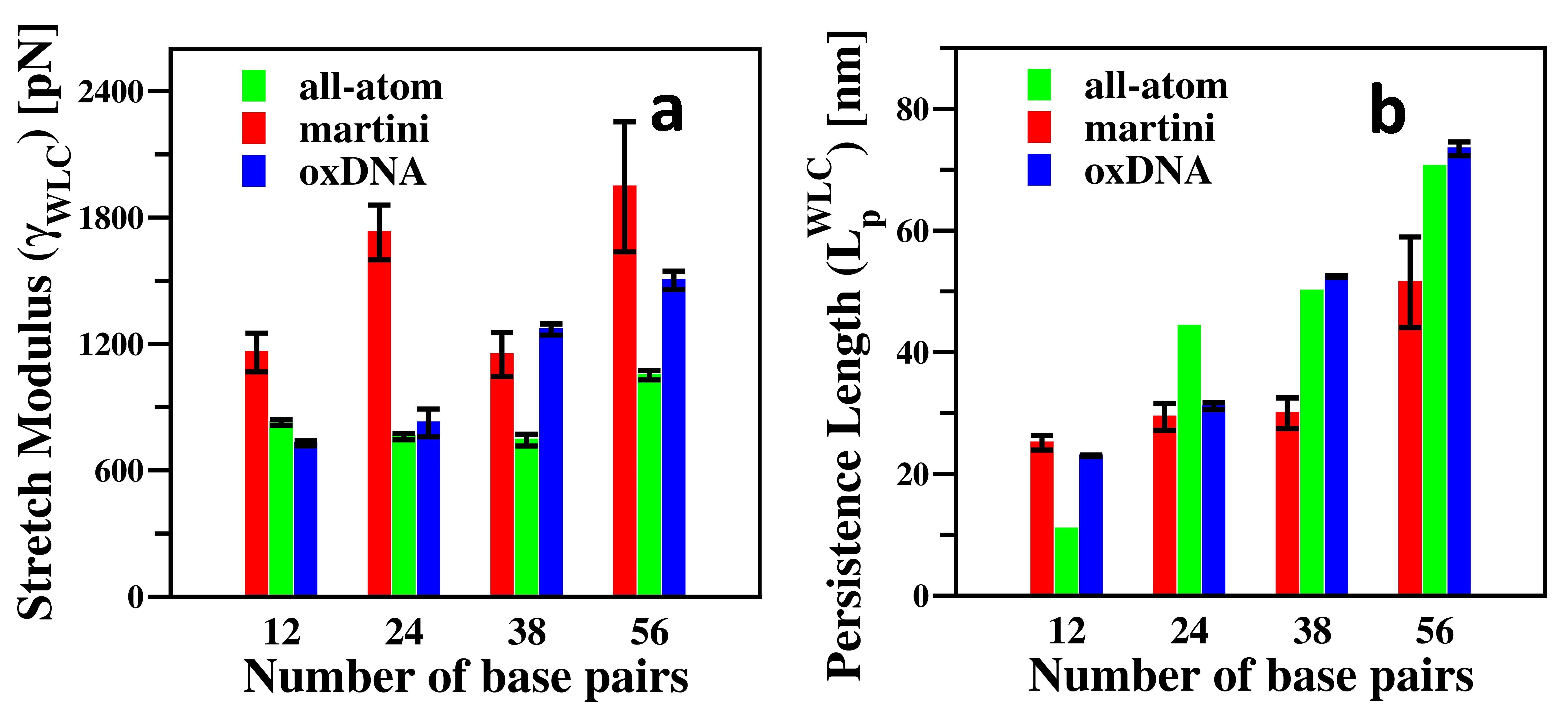}
 \caption{Stretch modulus and persistence length of CG dsDNA having different lengths. The values are obtained from the WLC model (equation \ref{7} and \ref{8}). }
 \label{fig3}
\end{figure}

We have also calculated the persistence length and the stretch modulus form the WLC model (equation \ref{7}) and elastic rod theory (equation \ref{8}) respectively. In order to calculate the persistence length using equation \ref{7}, we first calculate the end-to-end distance of the dsDNA and then numerically solved equation \ref{7} with the persistence length $L_p^{WLC}$ as a fitting parameter. The $L_p^{WLC}$ is then used in equation \ref{8} to calculate the stretch modulus ($\gamma^{WLC}$). Both the persistence length ($L_p^{WLC}$) and the stretch modulus obtained from the WLC model are in close agreement with the values  obtained from the all-atom MD simulations (see figure \ref{fig3} and table \ref{cgtab1}). \par

Also, the results from the WLC model are supporting a series of experiments and theoretical results reporting that small DNAs of few tens of bps have very high flexibility than those of kilo bps\cite{43, 80, 85, 89, 100}. Wiggins et al. using AFM study and WLC model reported  that the bending of short DNAs is larger than that of long DNAs\cite{85}. In another study, Yuan et al. using FRET and SAXS showed that DNAs with length in the range of 15-89 bps have higher flexibility than kilo bps DNAs\cite{89}. They obtained radius of gyration ($R_g$) from the scattering data\cite{89} and then $L_P$ has been obtained by numerically solving the following equation \ref{10},

\begin{equation}
\langle\ R_g^2\ \rangle=\frac{LL_P}{3}-L_P^2+\frac{2L_P^3}{L}-\frac{2L_P^4}{L^2}\left[1-exp\left(-\frac{L}{L_P}\right)\right] \label{10}
\end{equation}

The values of $L_P$ obtained in our coarse grained simulation are very close to the values obtained in the experiments\cite{89}. In a recent all-atom MD simulation study, Wu et al. reported similar results for small dsDNA\cite{86}. Based on the simulation results, they provided an empirical formula of persistence length ($L_p$) as a function of DNA bp length ($N$) which follows,

\begin{equation}
 L_P\left(N\right)=L_P^0-\frac{A}{\left(B+N\right)} \label{11}
\end{equation}
\begin{figure}[htbp]
 \centering
 \includegraphics[width=6.25in,keepaspectratio=true]{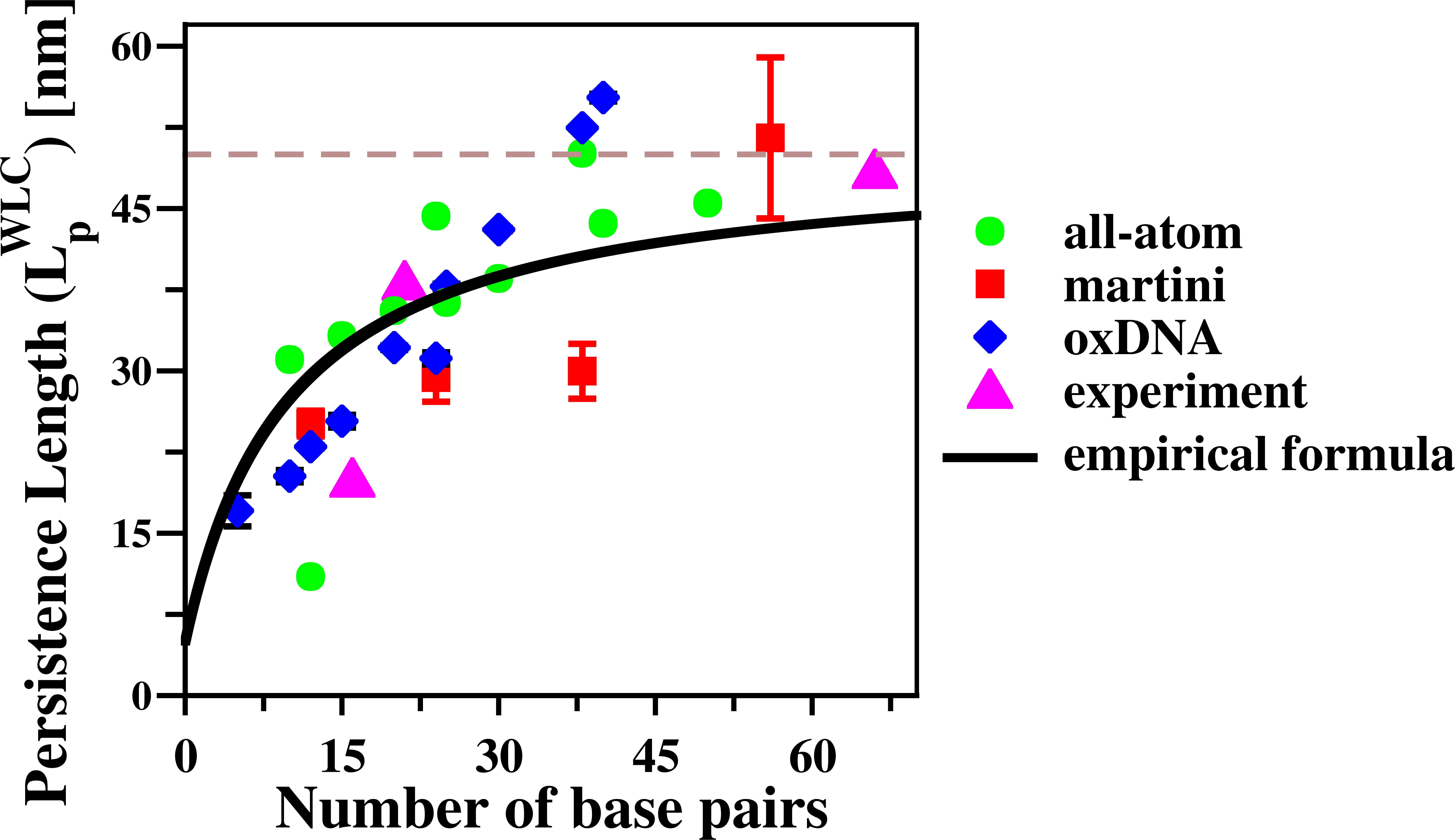}
 \caption{Length scale-dependent dsDNA flexibility. The persistence length of the all-atom and CG dsDNA is obtained from the WLC model. The experimental persistence length is obtained from the experimental of the short DNAs by Yuan et al.\cite{89}. The dashed line represents the persistence length of long dsDNA ~ 50 nm. The solid line is the empirical formula given by equation \ref{11}.The all-atom data are adapted from the reference \cite{17,86}}
 \label{fig4}
\end{figure}
where $L_P^0$ is the persistence length ($\sim$ 50 nm) of kilo base pair long DNA and $N$ is the number of bps. $A$ and $B$ are the fitting parameters and have values 450 nm and 10 respectively. In figure \ref{fig4} we have plotted equation \ref{11} as a function of $N$ along with the experimental results by Yuan et al.\cite{89}, all atom results by Garai et al.\cite{17} and Wu et al.\cite{86} and our simulation results. In order to compare better, we have also plotted few more data of oxDNA which has same lengths as done by Wu et al.\cite{86}.
We find that the $L_P$ calculated from all-atom and soft-Martini models are in close agreement with the experiment as well as with empirical equation \ref{11}. However, oxDNA as shown in blue diamond matches reasonably well with empirical formula and experimental for dsDNA lengths up to 30 bps.\par

To understand how salt concentration might affect the mechanical properties of the CG dsDNA we have carried out simulations at two different salt concentrations – 150 mM and 250 mM. Being a highly charged macromolecule, salt is known to have a strong effect on the structural and mechanical properties of DNA\cite{8, 17, 51, 84}. 
\begin{figure}[b]
 \centering
 \includegraphics[width=6.45 in,keepaspectratio=true]{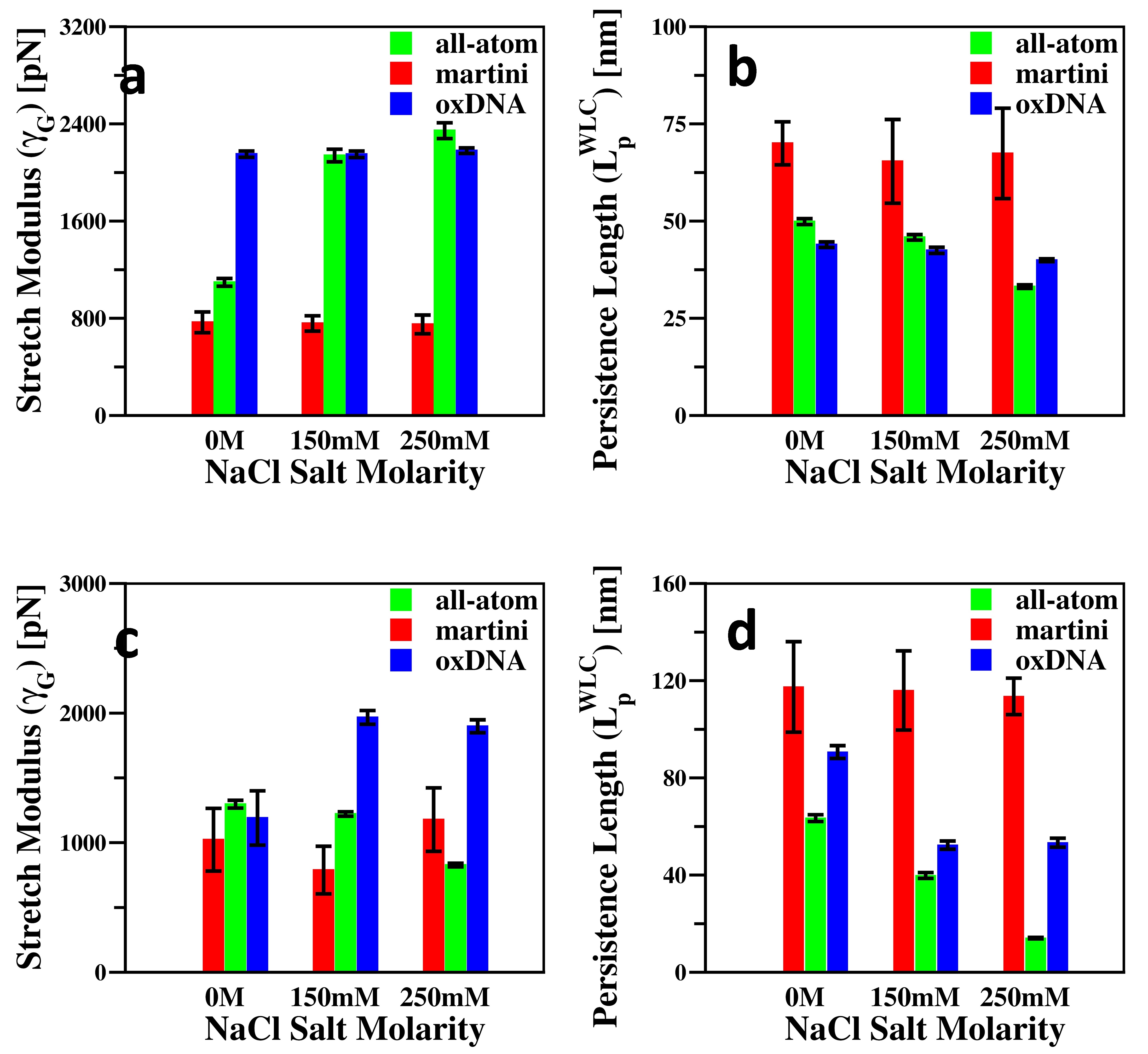}
 \caption{Salt dependent mechanical properties of dsDNA. (a) Stretch modulus and (b) persistence length of a 12 bps dsDNA in three different salt concentrations. (c) Stretch modulus and (d) persistence length of a 56 bps dsDNA in three different salt concentrations. Salt dependent stretch modulus and persistence length of other lengths of dsDNA has been provided in the supporting information.}
 \label{fig5}
\end{figure}
According to the Odijk-Skolnick-Fixman (OSF) model the persistence length decreases with ionic strength (I) as ~\ $I^{-1}$\cite{54, 71}. High salt concentration increases the screening of DNA backbone charges, which leads to more flexible dsDNA. Indeed, Garai et al. also concluded a similar trend in their all-atom simulation results\cite{17}. Several other studies also established the above results\cite{8, 21}. But in our Martini CG simulation results, we do not find any such trend. Both the stretch modulus and the persistence lengths almost remained invariant with increasing salt concentration for all the dsDNA lengths studied in this work (see figure \ref{fig5} and table S2 of SI). $\gamma$ and $L_P$ calculated by various methods as has been outlined above showing similar trends. Our simulation results indicate that the mechanics of coarse-grain Martini dsDNA does not depend on the salt concentration. The original Martini force-field paper by Uusitalo et al. also reported that the $L_P$ values changed from 68 $\pm$ 14 nm to 78 $\pm$ 11 nm as the NaCl concentration changed from 100mM to 1000mM\cite{78}. Within the error bar there is no change in the value of persistence length in going from 100 mM to 1000 mM. We think the less insensitivity of the CG Martini DNA mechanics with salt is because of the large CG beads and the way charge is distributed within the beads. In Martini model, the -1e charge is distributed in a single bead and most of the Na$^+$ ions are tightly bound to it. Because of the large radius of the CG beads the Na$^+$ ions are not able to penetrate into the DNA grooves and screen the backbone repulsion effectively. As a result, we see no change in the DNA mechanics with increasing salt concentration. In contrast, as reported in the literature, oXDNA model captures the effect of salt concentration on DNA mechanics very well\cite{75}. In the original  oxDNA paper, Snodin et al reported that the persistence length changes as the salt concentration varied from 100mM to 1000mM\cite{75}. We observe that for the oxDNA model the persistence length decreases as the salt concentration is increased (see figure \ref{fig5} and table S4 of SI). The effect of salt concentration becomes more prominent as the length of the dsDNA increases. Also, the stretch modulus increases as the salt concentration increases which is in agreement with many previous theoretical and experimental studies. Adding more salts screen the electrostatic interaction and help decrease the backbone repulsion leading to less fluctuation in the axial direction of the dsDNA. As a result, we observe an increment in the stretch modulus. \par

We have also performed few simulations of Martini dsDNA with soft elastic network (of lengths 12 bps, 24 bps, 38 bps) using polarizable water (PW) model. The polarizable CG water is made of three beads instead of one as in the standard Martini CG water model\cite{polar}. The central bead of the PW interacts by LJ potential whereas other two beads which carry a positive and negative charge interacts via a Coulombic interaction only, and lack any LJ interactions. We found that the dsDNA exhibits almost similar structural deviations in the PW bath. The stretch modulus values for 12 bps, 24 bps, 38 bps dsDNA in PW water are found to be around 981 $\pm$ 238, 900 $\pm$ 130, 818 $\pm$ 176 pN respectively. These values are very similar to the of values of 767 $\pm$ 85, 916 $\pm$ 160, 854 $\pm$ 75 pN in standard Martini CG water for 12 bps, 24 bps, 38 bps dsDNA respectively. Similarly, the persistence lengths are 41 $\pm$ 6, 43 $\pm$ 6, 57 $\pm$ 4 nm for 12 bps, 24 bps, 38 bps dsDNA respectively when simulated in PW water.

\begin{figure}[b]
 \centering
 \includegraphics[width=6.25in,keepaspectratio=true]{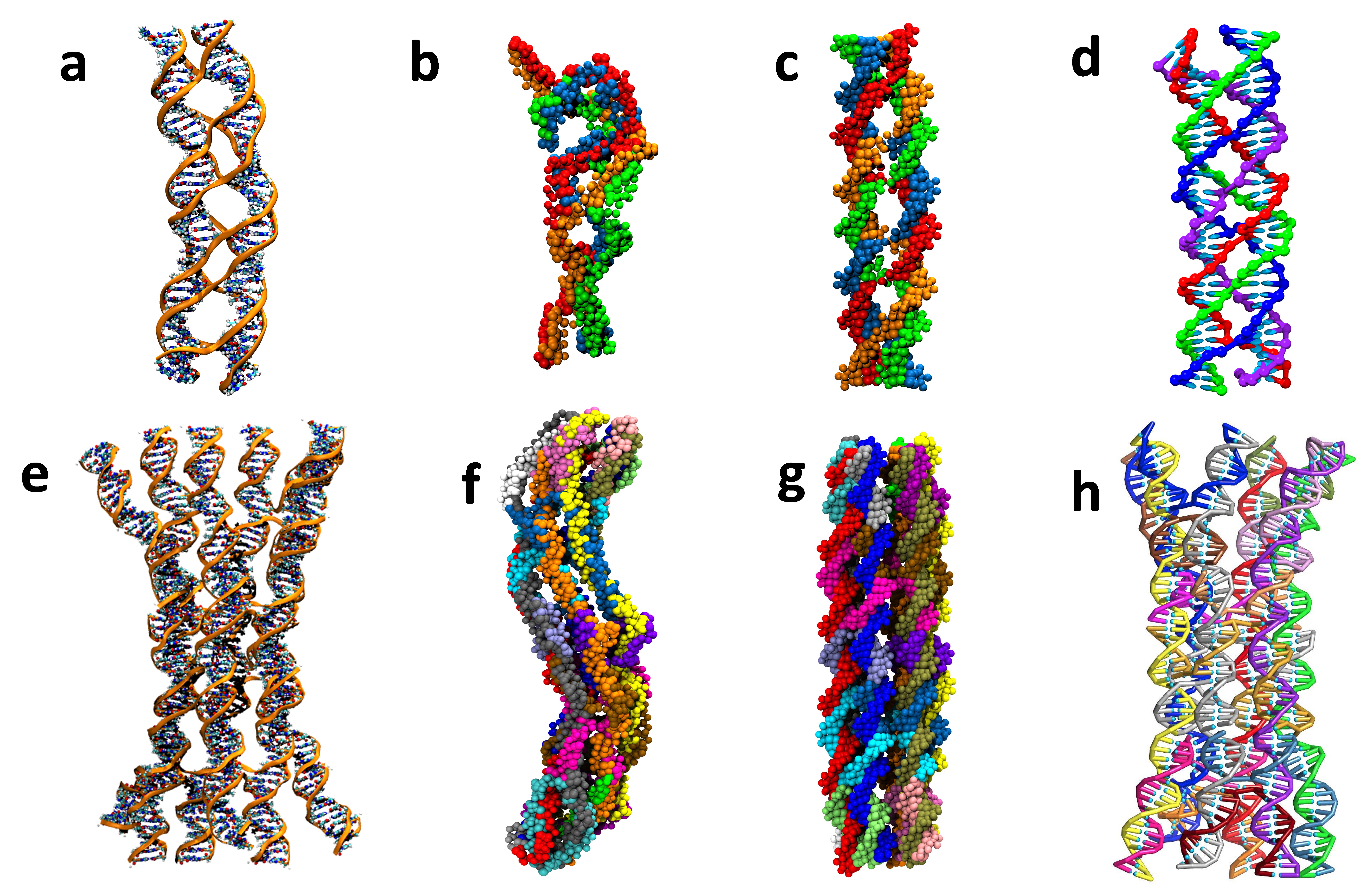}
 \caption{Instantaneous snapshots of DNA nanostructures. Final structure of paranemic crossover DNA (PX DNA) and 6-helix DNA nanotube (DNT) after long MD simulation. Final structure of PX DNA simulated using (a) all-atom (b) CG soft-Martini (c) CG stiff-Martini and (d) CG oxDNA model. Final structure of DNT simulated using (e) all-atom (f) CG soft-Martini (g) CG stiff-Martini and (h) CG oxDNA model.}
 \label{fig6}
\end{figure}
\subsection{Mechanical properties of DNA nanostructures}
Before evaluating the mechanical properties of DNA nanostructures like PX DNA or DNTs , we need to ensure that the time evolved CG structures of these nanostructures are stable and maintain integrity close to the experimental/all-atom structures. We observe that DNA nanostructures modeled with stiff elastic network in Martini framework is very rigid and has very little fluctuation over $\mu$s long simulations. In contrast, DNA nanostructures modeled with soft elastic network in Martini framework deformed and twisted very much and give rise to unrealistic structure. The time evolved oxDNA DNT reproduces reasonable structure and compare well with the all-atom MD simulated structure (see figure \ref{fig6}).\par

We have also computed the stretch modulus of PX DNA nanostructures using the elastic rod model. In order to calculate the contour length of the PX DNA, we divided the PX DNA into slices along the long axis, taking one base pair from each of the dsDNA. Then we add the center of mass (COM) distance of consecutive slices to get the contour length. The stretch modulus is then calculated using equation \ref{2}. The stretch modulus of PX-DNA is found to be 1548 $\pm$ 112 pN, 883 $\pm$ 312.21 pN, and 1758 $\pm$ 153 pN  for all-atom, soft-Martini and oxDNA model respectively. Previously, Santosh et al., by implementing non-equilibrium steered MD found that the stretch modulus of PX DNA is in the range of 1300-2300 pN\cite{98}. The all-atom and oxDNA model reproduces the results well within those range. Due to high structural fluctuation soft-Martini model underestimates the stretch modulus.\par

\begin{figure}[b]
 \centering
 \includegraphics[width=4.25 in,keepaspectratio=true]{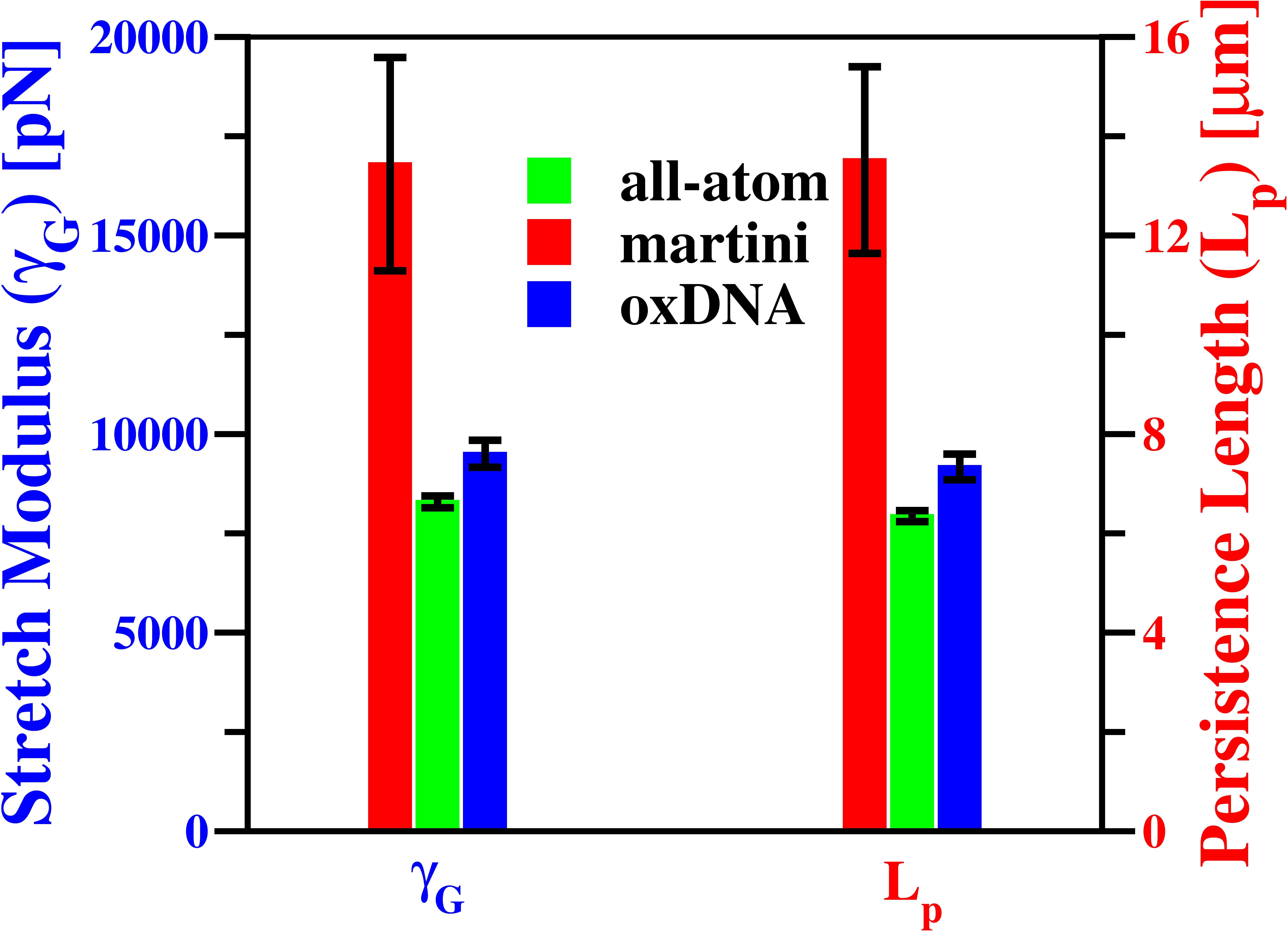}
 \caption{Stretch modulus and persistence length of CG DNA nanotube. The quantities are calculated using equation \ref{3} and \ref{9}.}
 \label{fig7}
\end{figure}

To calculate the stretch modulus and persistence length of DNT, we employed the same methodologies described earlier to estimate the mechanical properties of  DNA. First, we determine the contour length of the DNT at each time step. The length of each dsDNA of the constituent DNT is calculated by adding the COM distance of consecutive bps and then averaged over the six helices to get the contour length. Similar methodologies have been successfully implemented by Maiti and coworkers in their previous studies on DNTs\cite{25, 49}. We find that the stretch modulus of DNT modeled with oxDNA is closer to the all-atom and experimental values than the soft-Martini DNT (see figure \ref{fig7} and table \ref{cgtab1}). This is attributed to the fact that the soft-Martini DNA has a higher stretch modulus than the all-atom DNA which enhances the rigidity of the DNT. The oxDNA modeled DNT reproduces the stretch modulus value more accurately.  The calculated stretch modulus is then used in equation \ref{10} to calculate the persistence lengths of the DNTs. Calculated persistence lengths show similar trend as stretch modulus where soft-Martini model predicts slightly higher value compared to the oxDNA which is closer to the all-atom and experimental values\cite{24, 25, 49, 83}. As mentioned earlier, the stiff-Martini model of DNT has hardly any fluctuation and calculated stretch modulus and persistence length values are almost double than the experimental and all-atom values (see table S5). \par
\begin{figure}[htbp]
 \centering
 \includegraphics[width=6.25 in,keepaspectratio=true]{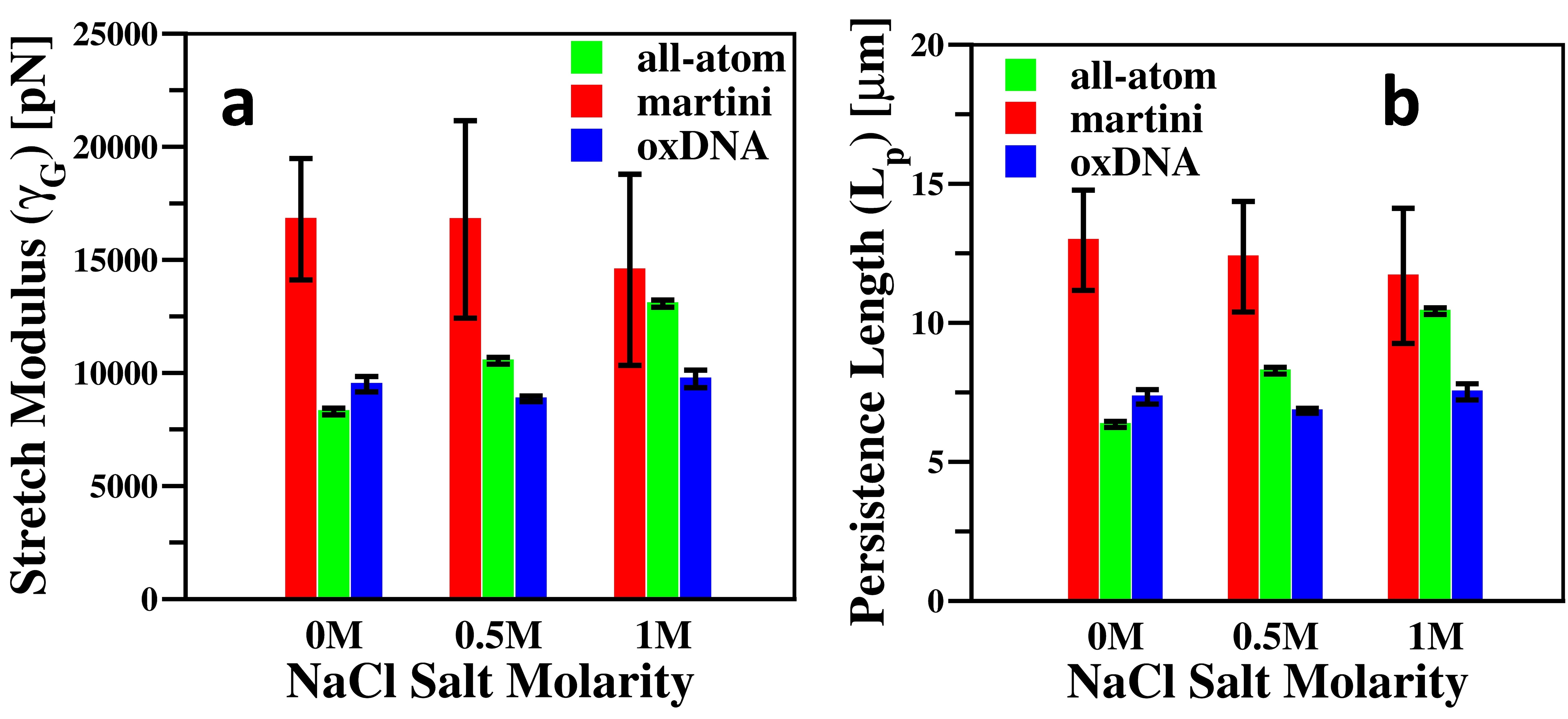}
 \caption{(a) Stretch modulus and (b) persistence lengths of CG DNA nanotubes (DNTs) at different salt concentrations.}
 \label{fig8}
\end{figure}

Presence of salt is known to have a large impact on the structure and mechanics of DNA nanostructures. Earlier, Naskar et al. have shown that the structural stability and mechanical rigidity of DNT can be increased by increasing monovalent salt concentration\cite{49}. Some recent experimental findings also reported that the charge screening by the MgCl$_2$ salt reconfigures the structure and mechanics of DNA nanotubes\cite{93,94}.  But, in our coarse-grain simulation study, we find the mechanics of DNT remains almost unchanged with salt concentration variation (see figure \ref{fig8}).\par

\begin{table}[ht!]
\centering
\caption{Mechanical properties of DNA and DNA nanostructures from this study as well as several previous studies.}
\label{cgtab1}
\resizebox{\textwidth}{!}{%
\begin{tabular}{llllll}
\hline \hline
System             & Lengths & Reference              & Model        & Stretch Modulus (pN)   & Persistence Length (nm) \\
\hline \hline \\
                   &         &                        &              &                        &                         \\
                   & 12 bp   &                        &              & 767 ($\gamma _G$), 1160 ($\gamma _{WLC}$)  & 70 ($L_P$), 25 ($L^{WLC}_P$)     \\
DNA                & 24 bp   & This work              & Soft-Martini & 916 ($\gamma _G$), 1730 ($\gamma _{WLC}$)  & 104 ($L_P$), 29 ($L^{WLC}_P$)    \\
                   & 38 bp   & (Theory)               &              & 855 ($\gamma _G$), 1150 ($\gamma _{WLC}$)  & 69 ($L_P$), 30 ($L^{WLC}_P$)     \\
                   & 56 bp   &                        &              & 1023($\gamma _G$), 1946 ($\gamma _{WLC}$)  & 117 ($L_P$), 51 ($L^{WLC}_P$)    \\
                   &         &                        &              &                        &                         \\
                   & 12 bp   &                        &              & 2152 ($\gamma _G$), 728 ($\gamma _{WLC}$)  & 44 ($L_P$), 23 ($L^{WLC}_P$)     \\
DNA                & 24 bp   & This work              & oxDNA        & 1470 ($\gamma _G$), 825 ($\gamma _{WLC}$)  & 50 ($L_P$), 31 ($L^{WLC}_P$)     \\
                   & 38 bp   & (Theory)               &              & 2210 ($\gamma _G$), 1269 ($\gamma _{WLC}$) & 77 ($L_P$), 52 ($L^{WLC}_P$)     \\
                   & 56 bp   &                        &              & 1398 ($\gamma _G$), 1502 ($\gamma _{WLC}$) & 91 ($L_P$), 73 ($L^{WLC}_P$)     \\
                   &         &                        &              &                        &                         \\
                   & 12 bp   &                        &              & 1096 ($\gamma _G$), 827 ($\gamma _{WLC}$)  & 50 ($L_P$), 11 ($L^{WLC}_P$)     \\
DNA                & 24 bp   & Garai et al.           & All-atom     & 871 ($\gamma _G$), 760 ($\gamma _{WLC}$)   & 46 ($L_P$), 44 ($L^{WLC}_P$)     \\
                   & 38 bp   & (Theory) \cite{17}     &              & 985 ($\gamma _G$), 744 ($\gamma _{WLC}$)   & 45 ($L_P$), 50 ($L^{WLC}_P$)     \\
                   & 56 bp   &                        &              & 1297 ($\gamma _G$), 1052 ($\gamma _{WLC}$) & 63 ($L_P$), 71 ($L^{WLC}_P$)     \\

                   &         &                        &              &                        &                         \\
                   & 16 bp   & Yuan et al.            &              &                        & 20                      \\
DNA                & 21 bp   & (experiment)           &              &                        & 38                      \\
                   & 66 bp   & \cite{89}              &              &                        & 48                      \\
                   &         &                        &              &                        &                         \\
                   &         &                        & All-atom     & 1548                   &                         \\
PX DNA             &         &                        & Soft-Martini & 883                    &                         \\
                   &         &                        & oxDNA        & 1758                   &                         \\
                   &         &                        &              &                        &                         \\
                   &         &                        &              &                        &                         \\
                   &         & Naskar et al. \cite{49}& All-atom     & 8295                   & 6350                    \\
DNA Nanotube (DNT) &         & This work              & Soft-Martini & 16800                  & 13520                   \\
                   &         & This work              & oxDNA        & 9505                   & 7340                   \\
                   &         & Wang et al.  \cite{89} & Experiment   &                        & 1000-2700                  
 \\ \hline \hline
\end{tabular}%
}
\end{table}

\newpage

\section{Conclusion}
In this work, we have investigated the mechanical properties of coarse-grained DNA and various DNA nanostructures including PX DNA and DNTs and compared those to the values obtained from fully atomistic simulations. We have employed two widely used CG models--Martini and oxDNA.  The lengths of the DNA studied were chosen to be short (ranging from 12 to 56 bps) which are biologically more important and can be simulated using fully atomistic description over longer time scale. We have employed worm-like chain model as well as the microscopic elastic rod theory to calculate all the mechanical properties. To calculate the stretch modulus and persistence length, we fit the contour length distribution, $P\left(L\right)$ and bending angle distribution, $P\left(\theta\right)$ to Gaussian respectively. We also computed the persistence length of dsDNA using end-to-end length distribution. We find that the mechanics of dsDNA and DNT estimated using the oxDNA and the Martini with soft-elastic network model, agree well with the experimental and all-atom calculations. In contrast, the stiff-elastic network model of Martini gives order of magnitude higher values of the stretch modulus and persistence lengths for the both dsDNA and DNT. Understanding of dsDNA or DNA nanostructure mechanics cannot be achieved using the stiff-elastic Martini model. Our results also indicate the length scale-dependent mechanical properties of dsDNA. From the WLC model, we find short dsDNA has a higher flexibility and lower persistence length than the long kilo bps dsDNA. Our results agree well with the several previous experimental and all-atom simulation results (see table \ref{cgtab1}). The empirical formula of length scale dependent persistence length substantiate the simulation results. We also find that oxDNA captures the structural deformation of DNA nanostructures more accurately than Martini. However, our calculation on these coarse-grain models of DNTs do not indicate any salt concentration-dependent mechanical properties. Proper distribution of charge on the beads may improve the results. \par
While the primary endeavor of this study has been that of measuring mechanical properties of dsDNA and DNT, we conclude that these methodologies can be further applied to any DNA like cylindrical molecule or nanostructures. Furthermore, these CG models can be implemented to describe and understand several other biophysical processes such as dsDNA melting, force-extension of dsDNA and DNA nanostructures, etc. However, if a more atomically thorough description of a DNA is required, all-atom MD methods are likely to be most suitable. \par
\section{Supplementary information}
Details of the simulated systems, snapshots of the CG dsDNA, contour length distribution, bending angle distribution, the salt effect on the mechanical properties.
\section{Conflict of interest}
The authors declare no competing financial interest.
\section{Acknowledgement}
SN acknowledges SRF fellowship from CSIR, India. We thank TUE-CMS, IISc Bangalore for the computation time. 
\bibliographystyle{apsrev4-2}
\bibliography{ms}
\end{document}


\title{Supporting Information\\Mechanical properties of DNA and DNA nanostructures: comparison of atomistic, Martini and oxDNA}

\author{Supriyo Naskar}
\email{supriyo@iisc.ac.in}
\affiliation{Center for Condensed Matter Theory, Department of Physics, Indian Institute of Science, Bangalore 560012, India}
\author{Prabal K. Maiti}
\email{maiti@iisc.ac.in}
\affiliation{Center for Condensed Matter Theory, Department of Physics, Indian Institute of Science, Bangalore 560012, India}
\maketitle

\newpage

\section{S\lowercase{equence and structure of ds}DNA \lowercase{and} DNA \lowercase{nanotube} (DNT)}

\subsection{Sequence of dsDNA}

\begin{table}[h]
  \caption{Sequences of Different dsDNAs studied in our simulations}
  \label{tbl:1}
  \begin{tabular}{ll}
    \hline
    Systems & Sequence \\
    \hline
      12 bp dsDNA & d(CGCGAATTCGCG)$_2$ \\
      24 bp dsDNA & d(CGCGATTGCCTAACGGACAGGCAT)$_2$ \\
      38 bp dsDNA & d(GCCGCGAGGTGTCAGGGATTGCAGCCAGCATCTCGTCG)$_2$ \\
      56 bp dsDNA & d(CGCGATTGCCTAACGGACAGGCATAGACGTCT\\
      & \              \   ATGCCTGTCCGTTAGGCAATCGCG)$_2$ \\
    \hline
  \end{tabular}
\end{table}
\newpage

\subsection{Schematic Diagram of DNT}
\captionsetup[figure]{format=cancaption,labelformat=cancaptionlabel}
\begin{figure}[H]
 \centering
 \includegraphics[width=6.0 in,keepaspectratio=true]{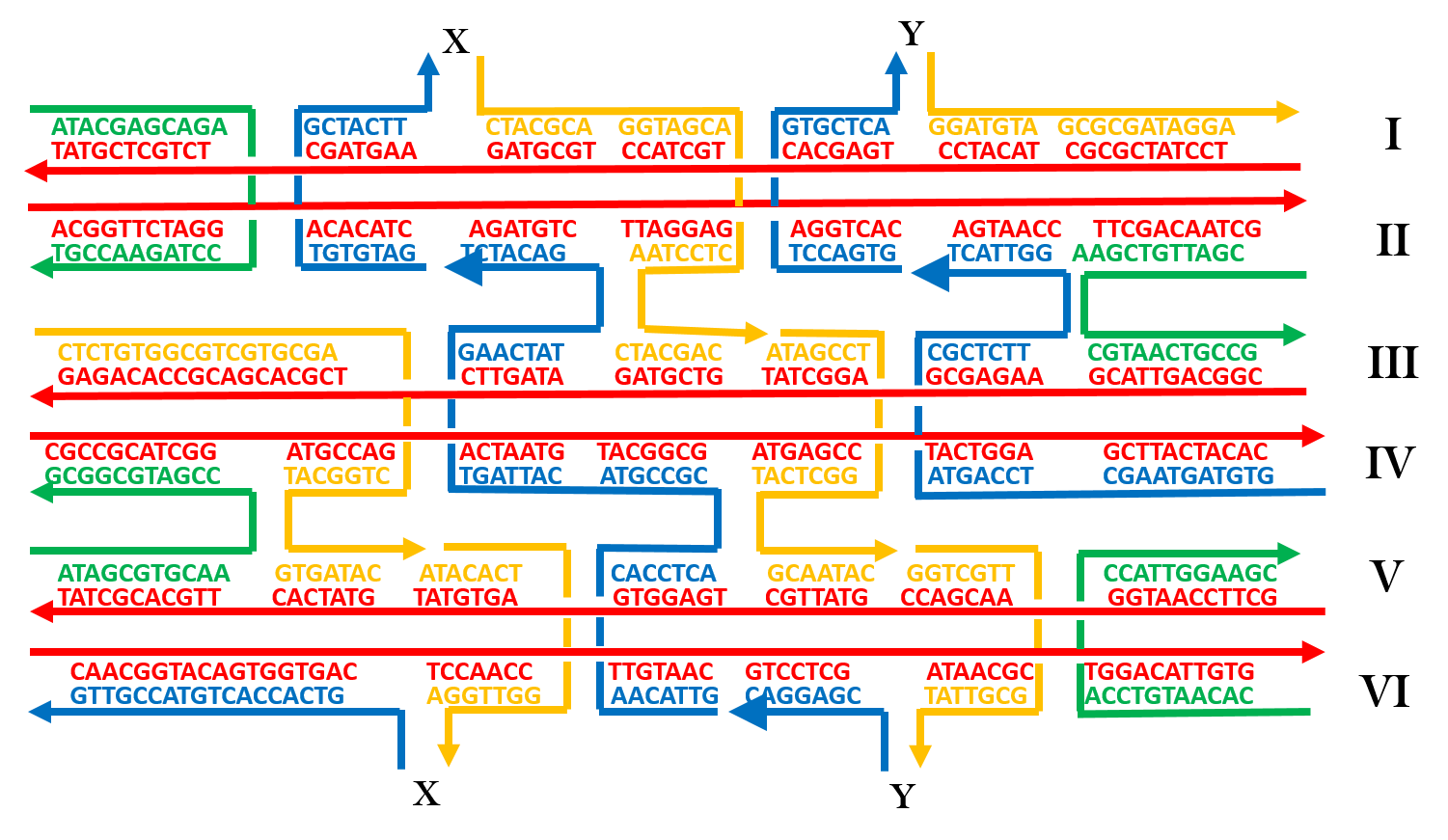}
 \caption{Schematic Diagram representing the DNT crossovers and sequence. The roman numbers indicating six different dsDNA strands. Different Colors represent different ssDNA strands. The arrows represent the polarity of the DNA from 5’ to 3’. The place X and Y are the position where the hexagonal bundle closes.}
 \label{fig1}
\end{figure}
\newpage

\section{F\lowercase{inal snapshots of the coarse-grained} (CG) \lowercase{ds}DNA}
\subsection{Soft-Martini}
\begin{figure}[H]
 \centering
 \includegraphics[width=6.25 in,keepaspectratio=true]{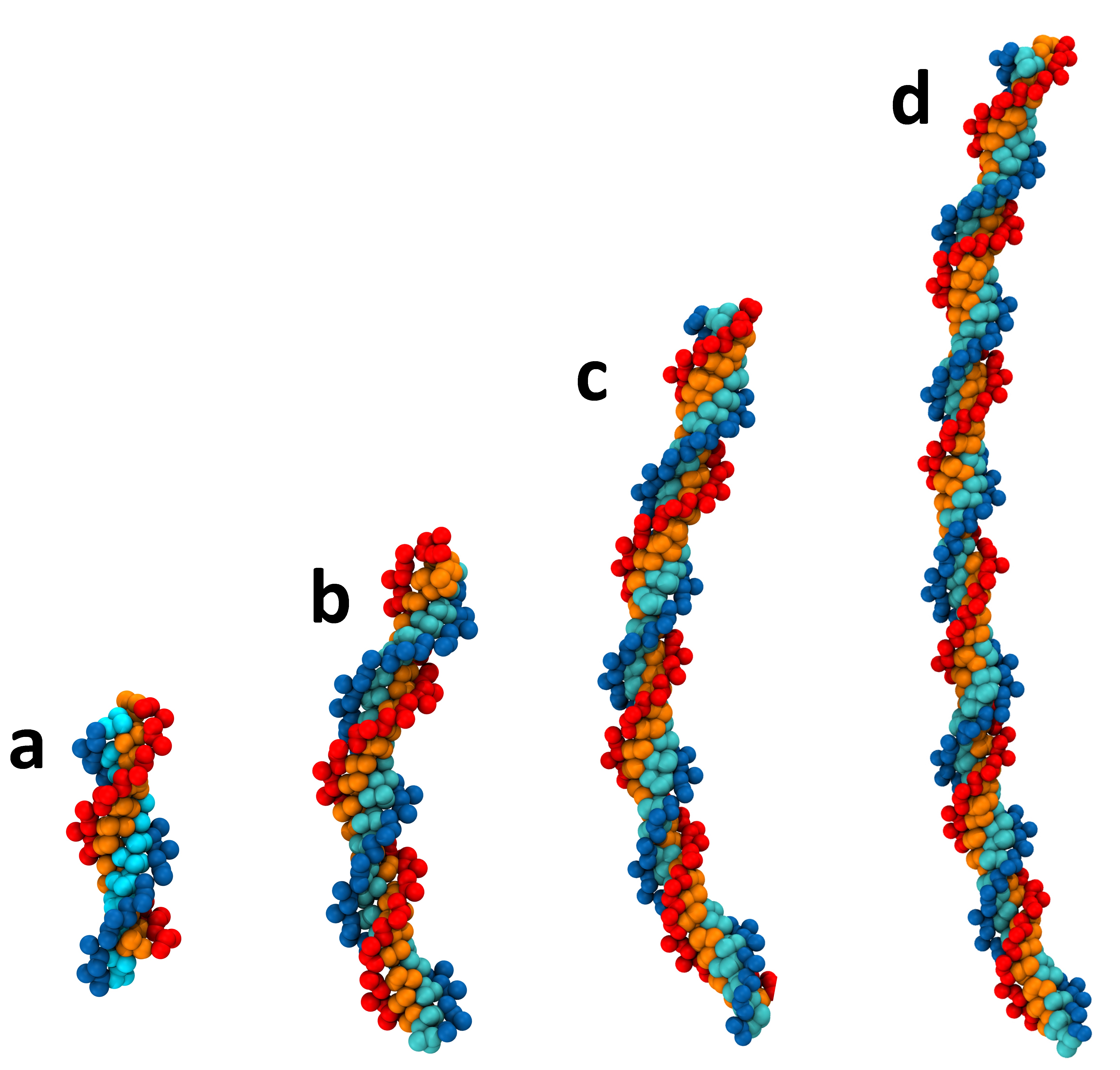}
 \caption{Final snapshots of the coarse-grained soft-martini dsDNA model after 2 $\mu$s long MD simulation. The length of the dsDNA is  (a) 12 bp (b)  24 bp (c) 38 bp (d) 56 bp.}
 \label{fig2}
\end{figure}
\newpage

\subsection{Stiff-Martini}
\begin{figure}[H]
 \centering
 \includegraphics[width=6.25 in,keepaspectratio=true]{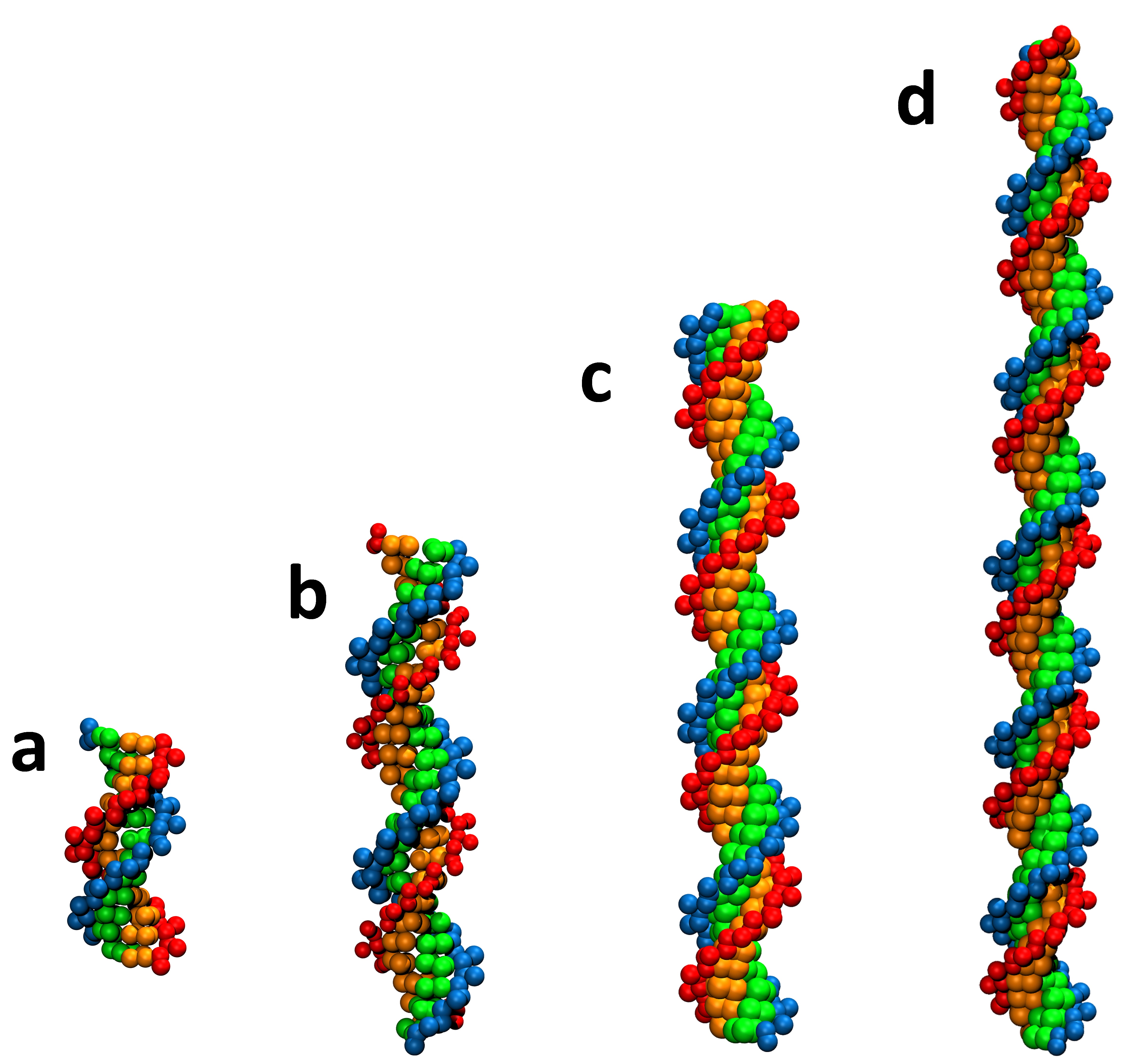}
 \caption{Final snapshots of the coarse-grained stiff-martini dsDNA model after 2 $\mu$s long MD simulation. The length of the dsDNA is  (a) 12 bp (b)  24 bp (c) 38 bp (d) 56 bp.}
 \label{fig3}
\end{figure}
\newpage

\subsection{OxDNA}
\begin{figure}[H]
 \centering
 \includegraphics[width=6.25 in,keepaspectratio=true]{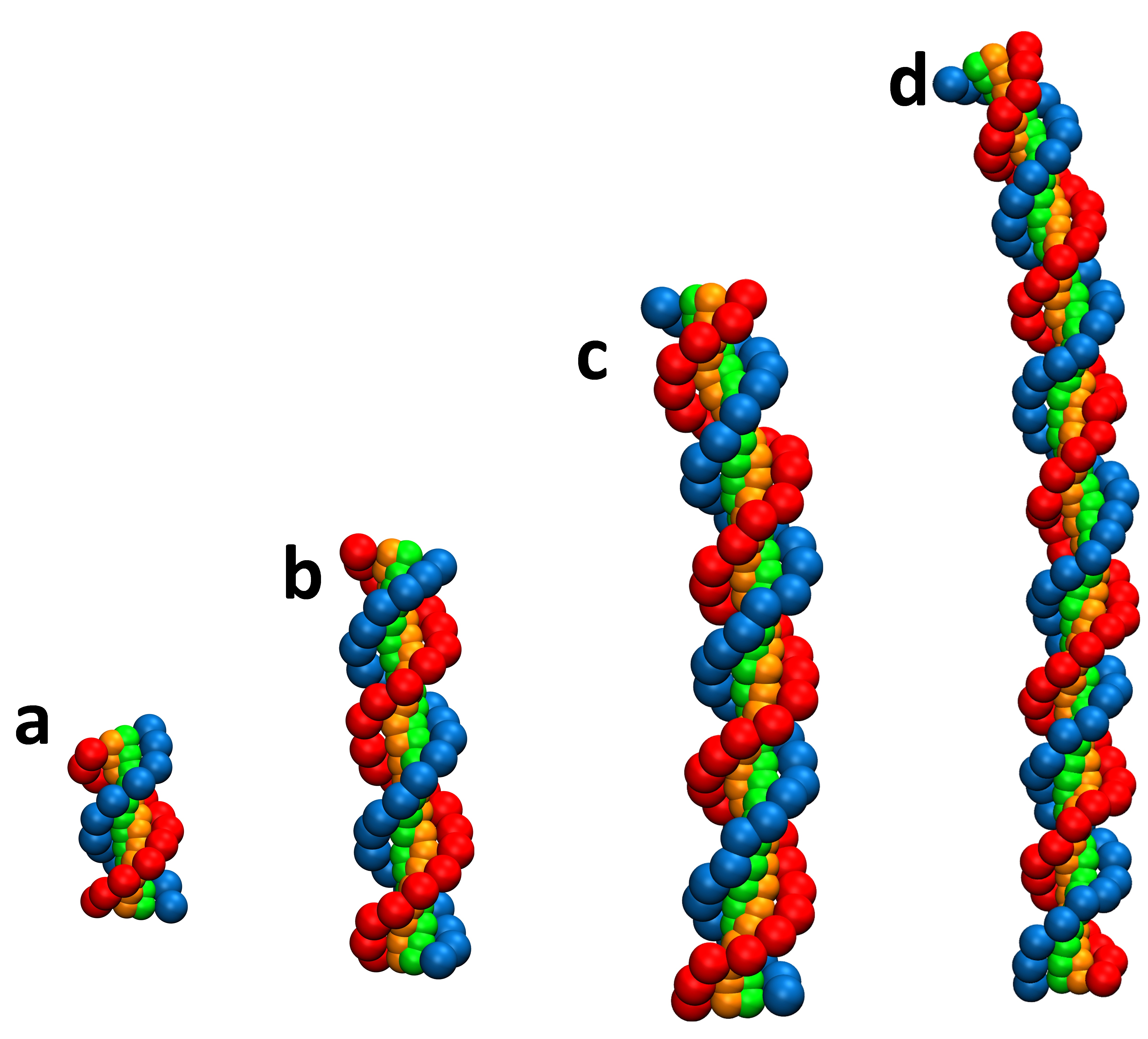}
 \caption{Final snapshots of the coarse-grained oxDNA dsDNA model after $2 \times 10^9$ long MD simulation steps. The length of the dsDNA is  (a) 12 bp (b)  24 bp (c) 38 bp (d) 56 bp. }
 \label{fig4}
\end{figure}
\newpage

\section{C\lowercase{ontour length and bending angle distribution of a 12bp ds}DNA}
\begin{figure}[H]
 \centering
 \includegraphics[width=4.25 in,keepaspectratio=true]{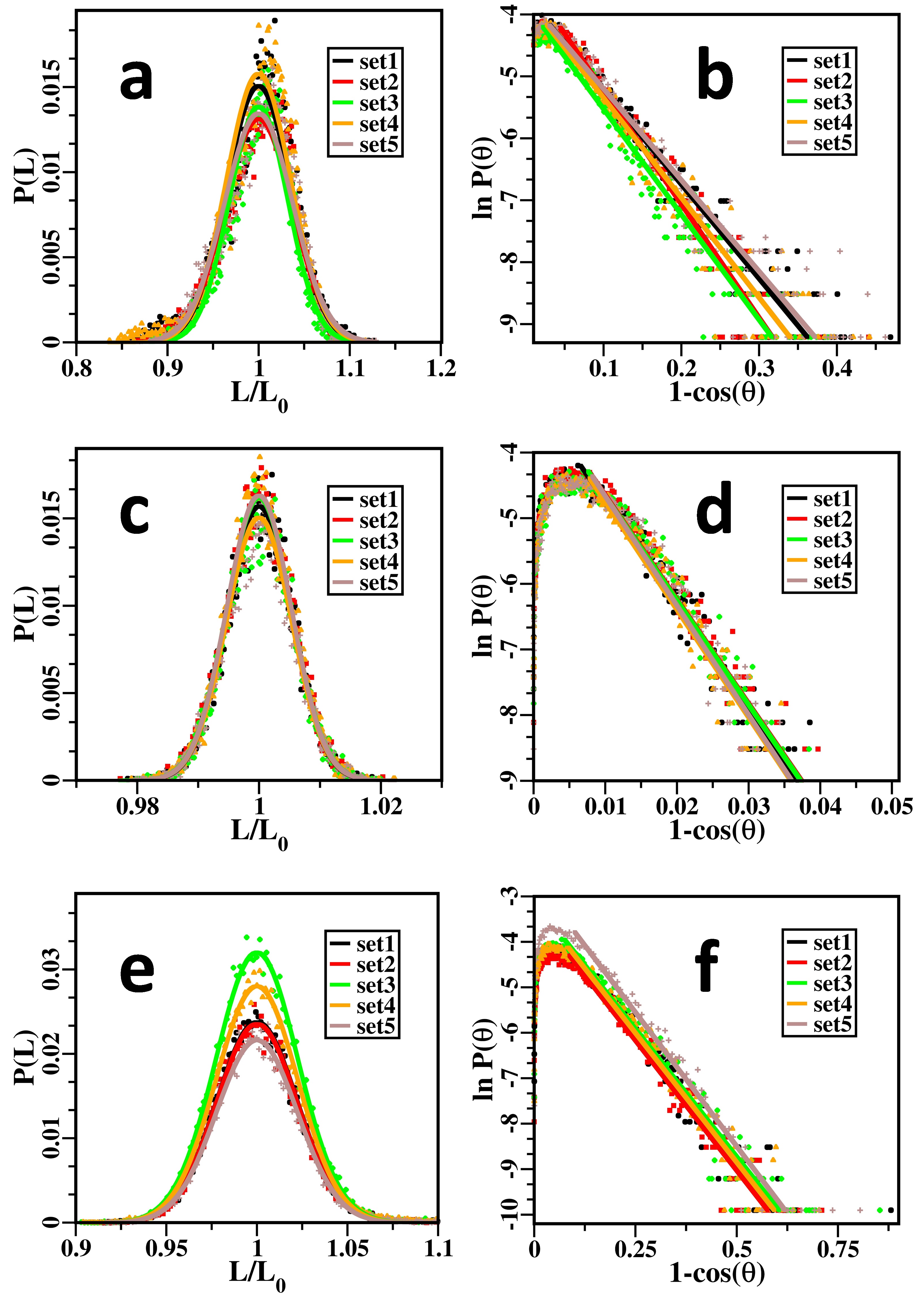}
 \caption{Contour length distribution and logarithm of bending angle distribution of a 12 bp dsDNA for various CG models. The trajectory is divided into five intervals of equal length and Different sets represent different time interval of the same trajectory. (a) Contour length distribution and (b) logarithm of bending angle distribution of a 12 bp soft martini dsDNA. (c) Contour length distribution and (d) logarithm of bending angle distribution of a 12 bp stiff martini dsDNA. (c) Contour length distribution and (d) logarithm of bending angle distribution of a 12 bp oxDNA.}
 \label{fig6}
\end{figure}
\newpage
\section{E\lowercase{lastic properties of }CG \lowercase{ds}DNA}
\begin{table}[ht]
\centering
\caption{Mechanical properties of soft-martini dsDNA at different salt concentration}
\label{tbl:2}
\resizebox{\textwidth}{!}{%
\begin{tabular}{lllllll}
System      & NaCl Salt Concentration & Stretch Modulus (pN) &                   &  & Persistence length (nm) &               \\
            &                         & $\gamma_G$               & $\gamma_{WLC}$        &  & $L_P$                       & $L_P^{WLC}$      \\
12 bp dsDNA & 0mM                     & 767.16  $\pm$ 85.29      & 1160. $\pm$ 91.69     &  & 70.03 $\pm$ 5.53            & 25.14 $\pm$ 1.19  \\
            & 150mM                   & 758.10 $\pm$ 63.43       & 1083.18 $\pm$ 178.42  &  & 65.38 $\pm$ 10.77           & 23.75 $\pm$ 2.00  \\
            & 250mM                   & 750.46 $\pm$ 77.14       & 1117.00  $\pm$ 192.69 &  & 67.42 $\pm$ 11.63           & 24.73 $\pm$ 2.04  \\
24 bp dsDNA & 0mM                     & 916.04 $\pm$ 159.80      & 1729.78 $\pm$ 130.23  &  & 104.40 $\pm$7.86            & 29.39 $\pm$ 2.23  \\
            & 150mM                   & 848.18 $\pm$ 409.60      & 973.97 $\pm$ 140.79   &  & 58.79 $\pm$ 8.50            & 16.94 $\pm$ 6.15  \\
            & 250mM                   & 1028.84 $\pm$  208.48    & 1563.22 $\pm$ 269.75  &  & 94.35 $\pm$ 16.28           & 38.09 $\pm$ 30.89 \\
38 bp dsDNA & 0mM                     & 854.68 $\pm$ 75.21       & 1150.11 $\pm$ 105.08  &  & 69.42 $\pm$ 6.34            & 29.98 $\pm$ 2.54  \\
            & 150mM                   & 1020.37 $\pm$ 117.91     & 1094.74 $\pm$ 241.57  &  & 66.07 $\pm$ 14.58           & 29.27 $\pm$ 1.91  \\
            & 250mM                   & 1225.32 $\pm$ 122.33     & 1083.33 $\pm$ 305.68  &  & 65.39 $\pm$ 18.45           & 29.59 $\pm$ 1.67  \\
56 bp dsDNA & 0mM                     & 1023.36 $\pm$ 241.80     & 1946.10 $\pm$ 308.86  &  & 117.46 $\pm$ 18.64          & 51.52 $\pm$ 7.44  \\
            & 150mM                   & 789.30 $\pm$ 183.54      & 1317.08 $\pm$ 635.32  &  & 116.00 $\pm$ 16.29          & 45.81 $\pm$ 9.73  \\
            & 250mM                   & 1178.63 $\pm$ 244.82     & 1881.61 $\pm$ 124.13  &  & 113.57 $\pm$ 7.49           & 55.70 $\pm$ 42.07
\end{tabular}%
}
\end{table}

\begin{table}[ht]
\centering
\caption{Mechanical properties of stiff-martini dsDNA at 0mM NaCl}
\label{tbl:3}
\resizebox{\textwidth}{!}{%
\begin{tabular}{lllll}
System & Stretch Modulus (pN) &                & Persistence length (nm) &             \\
       & $\gamma_G$           & $\gamma_{WLC}$ & $L_P$                   & $L_P^{WLC}$ \\
12 bp dsDNA & 38983.44 $\pm$ 789.02  & 9964.32 $\pm$ 204.59   & 601.42 $\pm$ 12.35   & 102.32 $\pm$ 0.82 \\
24 bp dsDNA & 47375.90 $\pm$ 2535.07 & 13920.78 $\pm$ 536.98  & 840.22 $\pm$ 32.41   & 200.77 $\pm$ 5.24 \\
       &                      &                &                         &             \\
38 bp dsDNA & 50697.57 $\pm$ 1162.17 & 9478.05 $\pm$ 248.87   & 572.07 $\pm$ 15.02   & 280.85 $\pm$ 1.41 \\
       &                      &                &                         &             \\
56 bp dsDNA & 51901.02 $\pm$ 416.46  & 18054.48 $\pm$ 2585.72 & 1089.72 $\pm$ 156.07 & 400.25 $\pm$ 7.69
\end{tabular}%
}
\end{table}
\newpage

\begin{table}[ht]
\centering
\caption{Mechanical properties of oxDNA at different salt concentration}
\label{tbl:4}
\resizebox{\textwidth}{!}{%
\begin{tabular}{llllll}
System & NaCl Salt Concentration & Stretch Modulus (pN) &                     & Persistence length (nm) &                   \\
       &                         & $\gamma_G$           & $\gamma_{WLC}$      & $L_P$                   & $L_P^{WLC}$       \\
12 bp dsDNA & 0mM                & 2151.60 $\pm$ 25.77  &  728.42 $\pm$ 11.91  & 43.96  $\pm$ 0.72  & 23.00 $\pm$ 0.14 \\
            & 150mM              & 2149.84 $\pm$ 26.43  &  704.18 $\pm$ 13.07  & 42.50  $\pm$ 0.79  & 22.61 $\pm$ 0.21  \\
            & 250mM              & 2180.13 $\pm$ 22.67  &  662.07 $\pm$ 6.15   & 39.96  $\pm$ 0.37  & 22.81 $\pm$ 1.74  \\
24 bp dsDNA & 0mM                & 1470.48 $\pm$ 74.31  &  825.73 $\pm$ 65.87  & 49.83  $\pm$ 3.98  & 31.16 $\pm$ 0.56  \\
            & 150mM              & 1520.40 $\pm$ 48.68  &  657.87 $\pm$ 37.86  & 39.71  $\pm$ 2.28  & 25.78 $\pm$ 0.30  \\
            & 250mM              & 1528.58 $\pm$ 56.27  &  649.91 $\pm$ 28.96  & 39.22  $\pm$ 1.75  & 25.42 $\pm$ 0.27  \\
38 bp dsDNA & 0mM                & 2209.62 $\pm$ 15.99  & 1268.88 $\pm$ 26.88  & 76.59  $\pm$ 1.62  & 52.45 $\pm$ 0.12 \\
            & 150mM              & 2288.43 $\pm$ 19.05  &  941.91 $\pm$ 21.16  & 56.85  $\pm$ 1.27  & 34.50 $\pm$ 0.14 \\
            & 250mM              & 2313.52 $\pm$ 23.13  &  909.22 $\pm$ 16.79  & 54.88  $\pm$ 1.01  & 33.73 $\pm$ 0.12  \\
56 bp dsDNA & 0mM                & 1398.29 $\pm$ 245.41 & 1502.17 $\pm$ 43.65  & 90.67  $\pm$ 2.63  & 73.45 $\pm$ 1.11 \\
            & 150mM              & 2309.68 $\pm$ 62.19  &  867.50 $\pm$ 28.31  & 52.36  $\pm$ 1.71  & 37.21 $\pm$ 0.67  \\
            & 250mM              & 2228.92 $\pm$ 58.90  &  884.10 $\pm$ 30.63  & 53.36  $\pm$ 1.85  & 36.78 $\pm$ 0.85 
\end{tabular}%
}
\end{table}
\newpage

\section{C\lowercase{ontour length distribution of }DNT}
\begin{figure}[H]
 \centering
 \includegraphics[width=6.25 in,keepaspectratio=true]{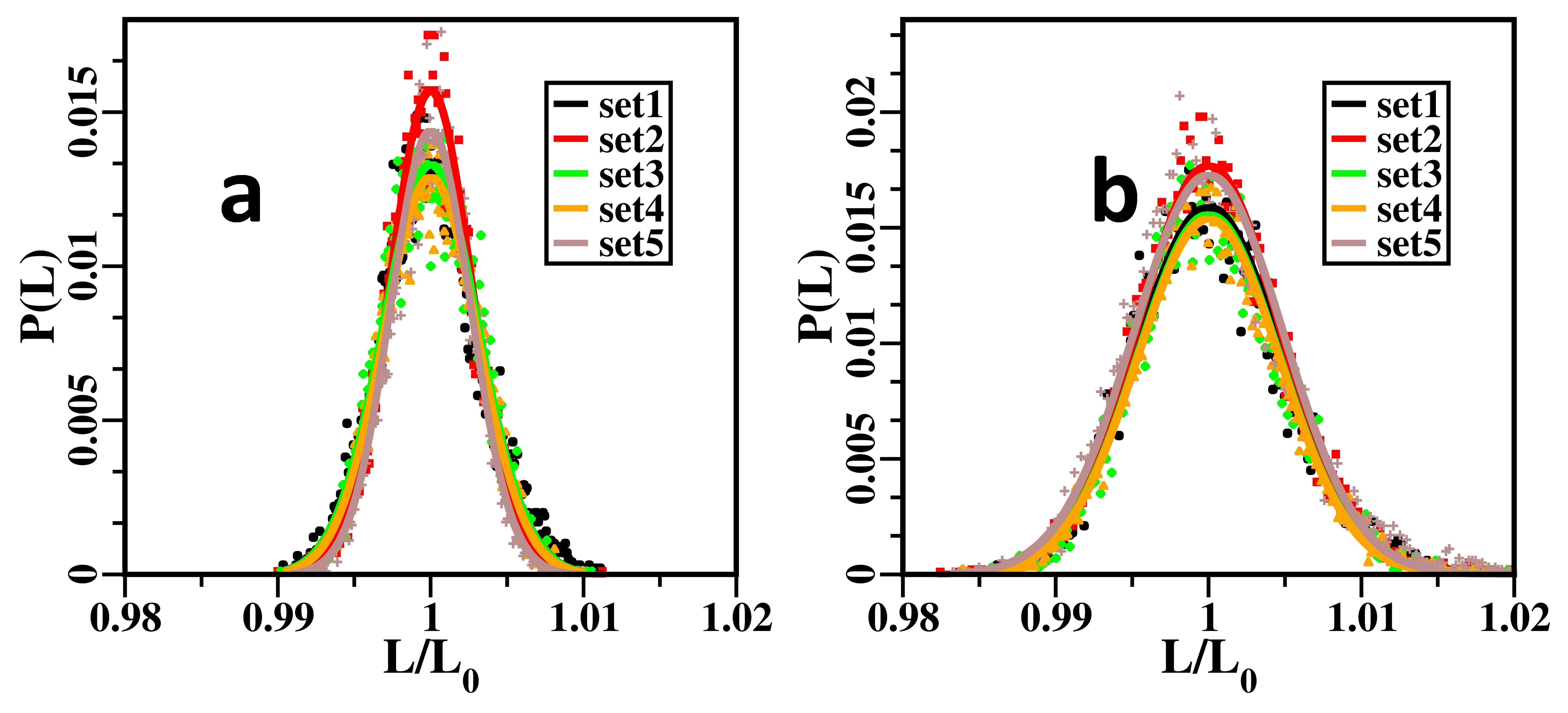}
 \caption{Contour length distribution of DNT for (a) soft martini (b) oxDNA.}
 \label{fig7}
\end{figure}
\newpage

\begin{figure}[H]
 \centering
 \includegraphics[width=3.125 in,keepaspectratio=true]{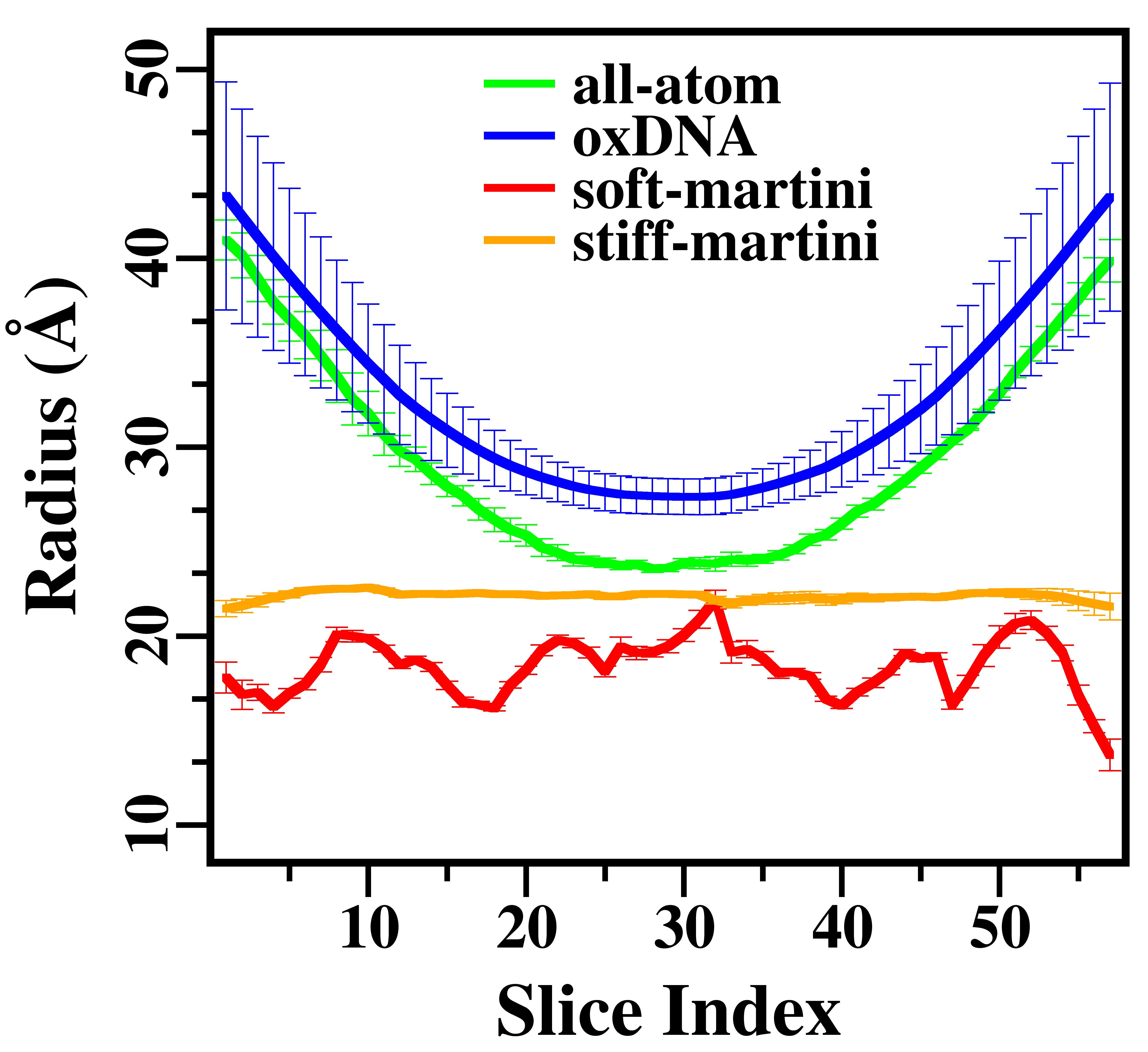}
 \caption{Radius of the pore of DNTs. The average value of the pore radius is used to estimate area moment of inertia, $I$ (equation 10 of the main airticle).}
 \label{fig8}
\end{figure}

\newpage
\begin{table}[ht]
\centering
\caption{Mechanical properties of DNT.}
\label{tbl:5}
\resizebox{\textwidth}{!}{%
\begin{tabular}{llllll}
Quantity & Salt concentration & all-atom             & soft martini            & stiff martini & oxDNA                \\
Stretch modulus (pN)        & 0 mM & 8294.87 $\pm$ 48.19 & 16799.39 $\pm$ 2684.95 & 113205.86 $\pm$ 3468.73 & 9504.72 $\pm$ 340.30 \\
         & 500 mM             & 10540.9 $\pm$ 148.13 & 16790.09  $\pm$ 4365.83 & --             & 8856.86 $\pm$ 121.76 \\
         & 1000 mM            & 13066.8 $\pm$ 155.91 & 14562.56 $\pm$ 4226.89  & --             & 9736.46 $\pm$ 386.64 \\
Persistence length ($\mu$m) & 0 mM & 6.35 $\pm$ 0.11     & 12.97 $\pm$ 1.8        & 88.55 $\pm$ 2.71        & 7.34 $\pm$ 0.26      \\
         & 500 mM             & 8.28 $\pm$ 0.12      & 12.38 $\pm$ 1.99        & --             & 6.84 $\pm$ 0.09       \\
         & 1000 mM            & 10.42 $\pm$ 0.12     & 11.69 $\pm$ 2.43        & --             & 7.52 $\pm$ 0.29     
\end{tabular}%
}
\end{table}